%% file: CLFpaper.tex
\documentclass{JINST}

\usepackage{amsmath}
\usepackage{amssymb}
\usepackage{caption}
\usepackage{graphics}
\usepackage{graphicx}
\usepackage{epsfig}
\usepackage{xspace}
\usepackage{afterpage}
\usepackage{textcomp}
\usepackage{setspace}
\usepackage{booktabs}
\usepackage{lineno}

\def\pao{Pierre Auger Observatory\xspace}
\def\mal{Malarg\"{u}e\xspace}

\def\degree{$^{\circ}$\xspace}

\newcommand{\cosec}{\operatorname{cosec}}

\title{Techniques for Measuring Aerosol Attenuation\\using the Central Laser Facility\\at the \pao}
\author{The Pierre Auger Collaboration\footnote{Authors are listed on the
following pages. E-mail: auger\_spokespersons@fnal.gov}}

\abstract
{
  The \pao in \mal, Argentina, is designed to study the properties of ultra-high
  energy cosmic rays with energies above $10^{18}~{\rm eV}$. It is a hybrid
  facility that employs a Fluorescence Detector to perform nearly calorimetric
  measurements of Extensive Air Shower energies.  To obtain reliable
  calorimetric information from the FD, the atmospheric conditions at the
  observatory need to be continuously monitored during data acquisition. In
  particular, light attenuation due to aerosols is an important atmospheric
  correction. The aerosol concentration is highly variable, so that the aerosol
  attenuation needs to be evaluated hourly. We use light from the Central Laser
  Facility, located near the center of the observatory site, having an optical
  signature comparable to that of the highest energy showers detected by the FD.
  This paper presents two procedures developed to retrieve the aerosol
  attenuation of fluorescence light from CLF laser shots.  Cross checks between
  the two methods demonstrate that results from both analyses are compatible,
  and that the uncertainties are well understood. The measurements of the
  aerosol attenuation provided by the two procedures are currently used at the
  \pao to reconstruct air shower data.
}


\begin{document}
\setlength{\topmargin}{3mm}

\newpage
\input{auger_collaboration_list_plus_MRoberts}
\newpage

\linenumbers

\input{intro}

\input{attenuation}

\input{clf}

\input{analysis}

\input{datanorm}

\input{lasersim}

\input{comparison}

\input{conclusions}

\nolinenumbers

\section*{Acknowledgments}
\input{acknowledgments}

\bibliographystyle{JHEP}
\bibliography{CLFpaper}

\clearpage

\end{document}

%% file: auger_collaboration_list_plus_MRoberts.tex
\par\noindent
{\bf The Pierre Auger Collaboration} \\
P.~Abreu$^{61}$, 
M.~Aglietta$^{49}$, 
M.~Ahlers$^{90}$, 
E.J.~Ahn$^{78}$, 
I.F.M.~Albuquerque$^{15}$, 
I.~Allekotte$^{1}$, 
J.~Allen$^{82}$, 
P.~Allison$^{84}$, 
A.~Almela$^{11,\: 7}$, 
J.~Alvarez Castillo$^{54}$
% ...

\par\noindent
{\it\footnotesize
$^{1}$ Centro At\'{o}mico Bariloche and Instituto Balseiro (CNEA-UNCuyo-CONICET), San 
Carlos de Bariloche, 
Argentina \\
% ...
$^{7}$ Instituto de Tecnolog\'{\i}as en Detecci\'{o}n y Astropart\'{\i}culas (CNEA, CONICET, UNSAM), 
Buenos Aires, 
Argentina \\
% ...
$^{11}$ Universidad Tecnol\'{o}gica Nacional - Facultad Regional Buenos Aires, Buenos Aires,
Argentina \\
% ...
$^{15}$ Universidade de S\~{a}o Paulo, Instituto de F\'{\i}sica, S\~{a}o Paulo, SP, 
Brazil \\
% ...
$^{49}$ Istituto di Fisica dello Spazio Interplanetario (INAF), Universit\`{a} di Torino and 
Sezione INFN, Torino, 
Italy \\
% ...
$^{54}$ Universidad Nacional Autonoma de Mexico, Mexico, D.F., 
Mexico \\
% ...
$^{61}$ LIP and Instituto Superior T\'{e}cnico, Technical University of Lisbon, 
Portugal \\
% ...
$^{78}$ Fermilab, Batavia, IL, 
USA \\
% ...
$^{82}$ New York University, New York, NY, 
USA \\
% ...
$^{84}$ Ohio State University, Columbus, OH, 
USA \\
% ...
$^{90}$ University of Wisconsin, Madison, WI, 
USA \\
% ...
\par\noindent
(\ddag) Deceased \\
(a) Now at Konan University \\
(b) Also at the Universidad Autonoma de Chiapas on leave of absence from Cinvestav \\
(c) Now at University of Maryland \\
(d) Now at NYU Abu Dhabi \\
(e) Now at Defence Science and Technology Organisation, Australia \\
}

%% file: intro.tex
\section{Introduction}
\label{sec:intro}

Direct measurements of primary cosmic rays at ultra-high energies (above
$10^{18}~{\rm eV}$) above the atmosphere are not feasible because of their
extremely low flux. The properties of primary particles --~energy, mass
composition, arrival direction~-- are deduced from the study of cascades of
secondary particles of Extensive Air Showers (EAS), originating from the
interaction of cosmic rays with air molecules. The \pao~\cite{auger} in
Argentina (mean altitude about 1400\,m a.s.l.) combines two well-established
techniques: the Surface Detector, used to measure photons and charged particles
produced in the shower at ground level; the Fluorescence Detector, used to
measure fluorescence light emitted by air molecules excited by secondary
particles during shower development. The Fluorescence Detector (FD)~\cite{FD}
consists of 24~telescopes located at four sites around the perimeter of the
Surface Detector (SD) array. It is only operated during clear nights with a low
illuminated moon fraction. The field of view of a single telescope is 30\degree
in azimuth, and 1.5\degree to 30\degree in elevation. Each FD site covers
180\degree in azimuth. The hybrid feature and the large area of 3000~km$^2$ of
the observatory enable the study of ultra-high energy cosmic rays with much
better precision and much greater statistics than any previous experiment.

The fluorescence technique to detect EAS makes use of the atmosphere as a giant
calorimeter whose properties must be continuously monitored to ensure a reliable
energy estimate. Atmospheric parameters influence both the production of
fluorescence light and its attenuation towards the FD telescopes. The molecular
and aerosol scattering processes that contribute to the overall attenuation of
light in the atmosphere can be treated separately. In particular, aerosol
attenuation of light is the largest time dependent correction applied during air
shower reconstruction, as aerosols are subject to significant variations on time
scales as little as one hour.  If the aerosol attenuation is not taken into
account, the shower energy reconstruction is biased by 8~to 25\% in the energy
range measured by the \pao~\cite{segev_atmo}. On average, 20\% of all showers
have an energy correction larger than 20\%, 7\% of showers are corrected by more
than 30\% and 3\% of showers are corrected by more than 40\%. Dedicated
instruments are used to monitor and measure the aerosol parameters of interest:
the aerosol extinction coefficient~$\alpha_{\rm aer}(h)$, the normalized
differential cross section --~or phase function~-- $P(\theta)$, and the
wavelength dependence of the aerosol scattering, parameterized by the ${\textrm
\AA}$ngstrom coefficient~$\gamma$.

\begin{figure}[htb]
  \begin{center}
    \includegraphics*[width=1.\textwidth,clip]{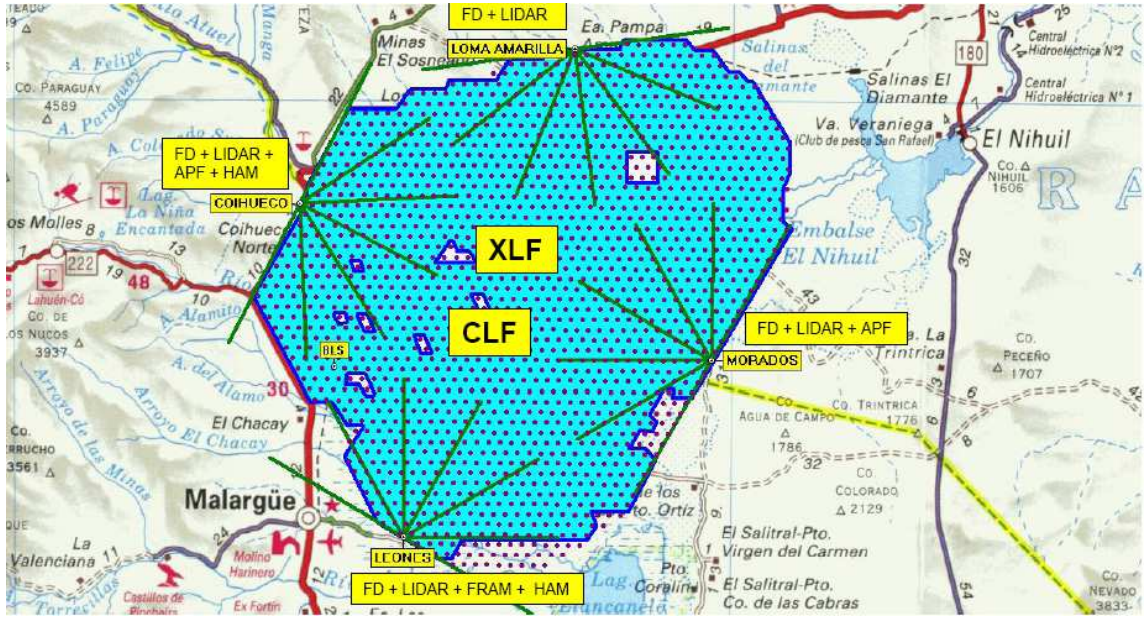}
    \caption{
      \label{fig:map_atmo}
      Map of the \pao in Argentina. Dots represent SD stations, which are
      separated by 1.5 km. The green lines represent the field of view of the six
      telescopes of each of the four fluorescence detectors at the periphery of
      the SD array. The position of the atmospheric monitoring devices is shown.
    }
  \end{center}
\end{figure}

At the \pao, molecular and aerosol scattering in the near UV are measured using
a collection of dedicated atmospheric monitors~\cite{segev_atmo}. One of these
is the Central Laser Facility (CLF)~\cite{clfjinst} positioned close to the
center of the array, as shown in Fig.~\ref{fig:map_atmo}. A newly built second
laser station, the eXtreme Laser Facility (XLF), positioned north of the CLF,
has been providing an additional test beam since 2009. The two systems produce
calibrated 355~nm vertical and inclined laser shots during FD data acquisition.
These laser facilities are used as test beams for various applications: to
calibrate the pointing direction of telescopes, for the determination of the
FD/SD time offset, and for measuring the vertical aerosol optical
depth~$\tau_{\rm aer}(h)$ and its differential~$\alpha_{\rm aer}(h)$. An hourly
aerosol characterization is provided in the FD field of view with two
independent approaches using the same CLF vertical laser events. In the near
future, those approaches will be applied to XLF vertical events. The FRAM
robotic telescope is used for a passive measurement of the total optical depth
of the atmosphere, the horizontal attenuation monitors (HAM) at two of the FD
sites are used to characterize the optical properties of the atmosphere close to
the ground.

In addition to the CLF and XLF, four monostatic LIDARs~\cite{lidar_paper} and
four Infrared Cloud Cameras~\cite{karim_icrc} --~one at each FD site~-- are
devoted to cloud and aerosol monitoring. During FD data acquisition, the LIDARs
continuously operate outside the FD field of view and detect clouds and aerosols
by analyzing the backscatter signal  of a 351 nm pulsed laser beam.  The cloud
cameras use passive measurements of the infrared light and provide a picture of
the field of view of every FD telescope every 5~minutes.

To measure the Aerosol Phase Function (APF), a Xenon flash lamp at two of the FD
sites fires a set of five shots with a repetition rate of 0.5~Hz once every
hour~\cite{apf}. The shots are fired horizontally across the field of view of
five out of the six telescopes in each building. The resulting angular
distribution of the signal gives the total scattering phase function $P(\theta)$
as a function of the scattering angle $\theta$.

In this paper, we will describe the analysis techniques used to estimate aerosol
attenuation from CLF laser shots. In Sec.~\ref{sec:att} we will review
atmospheric attenuation due to aerosols and molecules. In Sec.~\ref{sec:instr},
we will discuss the setup, operation and calibration of the CLF.
Sec.~\ref{sec:analysis} contains the description of the two analysis methods
used to estimate the aerosol attenuation. Comparisons between the two methods
and conclusions follow in Sec.~\ref{sec:comparison} and~\ref{sec:conclusions}.

%% file: attenuation.tex
\section{Atmospheric Attenuation}
\label{sec:att}

Molecules in the atmosphere predominantly scatter, rather than absorb,
fluorescence photons in the UV range\footnote{The most absorbing atmospheric
gases in the atmosphere are ozone and NO$_2$. In the 300 to 400~nm range, the
contribution of their absorption to the transmission function is
negligible~\cite{segev_atmo}.}.  Molecular and aerosol scattering processes can
be treated separately. In the following, the term ``attenuation'' is used to
indicate photons that are scattered in such a way that they do not contribute to
the light signal recorded by the FD. The molecular and aerosol attenuation
processes can be described in terms of atmospheric transmission coefficients
$T_{\rm mol}(\lambda,s)$ and $T_{\rm aer}(\lambda,s)$, indicating the fraction
of transmitted light intensity as a function of the wavelength~$\lambda$ and the
path length~$s$. The amount of fluorescence light recorded at the FD aperture
$I(\lambda,s)$ can be expressed in terms of the light intensity at the source
$I_0(\lambda,s)$ as
\begin{equation}
\label{eqn:I_lambda}
  I(\lambda,s) = I_0(\lambda,s) \cdot T_{\rm mol}(\lambda,s)
    \cdot T_{\rm aer}(\lambda,s)
    \cdot (1 + {\rm H.O.}) \cdot \frac{{\rm d}\Omega}{4\pi},
\end{equation}
where H.O.\ are higher order corrections due to multiple scattering and ${\rm
d}\Omega$ is the solid angle subtended by the telescope aperture as seen from
the light source.

An accurate measurement of the transmission factors during data acquisition is
necessary for a reliable reconstruction of the shower and for proper
measurements of the physical properties of the primary particle (energy, mass
composition, etc). While the molecular transmission factor $T_{\rm
mol}(\lambda,s)$ can be determined analytically once the vertical profiles of
atmospheric temperature, pressure, and humidity are known, the aerosol
transmission factor $T_{\rm aer}(\lambda,s)$ depends on the aerosol distribution
$n_{\rm aer}(r,h)$, where $r$ is the aerodynamic radius of the aerosols and $h$
is the height above the ground.

The molecular transmission factor $T_{\rm mol}(\lambda,s)$ is a function of the
total wavelength-dependent Rayleigh scattering cross section $\sigma_{\rm
mol}(\lambda)$ and of the density profile along the line of sight $s$ in
atmosphere $n_{\rm mol}(s)$,
\begin{equation}
\label{eqn:T_mol}
  T_{\rm mol}(\lambda,s) = \exp{ \left(-\int
    \sigma_{\rm mol}(\lambda) \, n_{\rm mol}(s) \, {\rm d} s \right) }.
\end{equation}
The Rayleigh scattering cross section $\sigma_{\rm mol}(\lambda)$ is
\begin{equation}
\label{eqn:sigma_mol}
  \sigma_{\rm mol}(\lambda) = \frac{24 \pi^3}{N_s^2 \lambda^4}
    \cdot \left(\frac{n_{\rm air}^2 - 1}{n_{\rm air}^2 + 2} \right)
    \cdot F_{\rm air}(\lambda),
\end{equation}
where $N_{\rm s}$ is the atmospheric molecular density, measured in molecules
per m$^{-3}$, $n_{\rm air}$ is the refractive index of the air, and $F_{\rm air}$
is the King factor that accounts for the anisotropy in the scattering introduced
by the non-spherical N$_2$, O$_2$ molecules~\cite{king_factor}.

The atmospheric density profile along the line of sight $n_{\rm mol}(s)$ is
calculated using altitude-dependent temperature and pressure profiles,
\begin{equation}
\label{eqn:N_mol}
    n_{\rm mol}(s) = \frac{N_{\rm A}}{R} \cdot \frac{p(h)}{T(h)},
\end{equation}
where $N_{\rm A}$ is Avogadro's number and $R$ is the universal gas constant.

Temperature, pressure and humidity vertical profiles of the atmosphere were
recorded from August 2002 to December 2010 by performing an intensive campaign
of radiosonde measurements above the site of the \pao~\cite{epj_atmo}. A set of
data was taken about every 20\,m during the ascent. The balloons were able to
reach altitudes of 25~km a.s.l.\ on average. Vertical profiles are complemented
by temperature, pressure and humidity data from five ground-based weather
stations. The measured profiles from these launches have been averaged to form
monthly mean profiles (\mal Monthly Models) which can be used in the simulation
and reconstruction of showers~\cite{epj_atmo,segev_atmo}. Currently, the Global
Data Assimilation System (GDAS) is used as a source for atmospheric profiles.
GDAS combines measurements and forecasts from numerical weather prediction to
provide data for the whole globe every three hours. For the location of the
\pao, reasonable data have been available since June 2005. Comparisons with
on-site measurements demonstrate the applicability of the data for air shower
analyses~\cite{gdas_paper}.

Aerosol scattering can be described by Mie scattering theory. However, it relies
on the assumption of spherical scatterers, a condition that is not always
fulfilled. Moreover, scattering depends on the nature of the particles.  A
program to measure the dimensions and nature of aerosols at the \pao is in
progress and already produced first results, but more study is
needed~\cite{maria_aerosol}.  Therefore, the knowledge of the aerosol
transmission factor $T_{\rm aer}(\lambda,s)$ depends on frequent field
measurements of the vertical aerosol optical depth $\tau_{\rm aer}(h)$, the
integral of the aerosol extinction $\alpha_{\rm aer}(z)$ from the ground to a
point at altitude $h$ observed at an elevation angle $\varphi_2$, assuming a
horizontally uniform aerosol distribution (cf.\ Fig.~\ref{fig:laser_geometry}),
\begin{equation}
\label{eqn:T_aer_alpha}
  T_{\rm aer}(\lambda,h)
    = \exp{ \left(-\int^h_0 \alpha_{\rm aer}(z) {\rm d} z/ \sin{\varphi_2} \right) }
    = \exp{ \left[- ( \tau_{\rm aer}(h) / \sin{\varphi_2}) \right] }.
\end{equation}
Hourly measurements of $\tau_{\rm aer}(h)$ are performed at each FD site using
the data collected from the CLF.

\begin{figure}[htb]
  \begin{center}
    \includegraphics*[width=0.6\textwidth,clip]{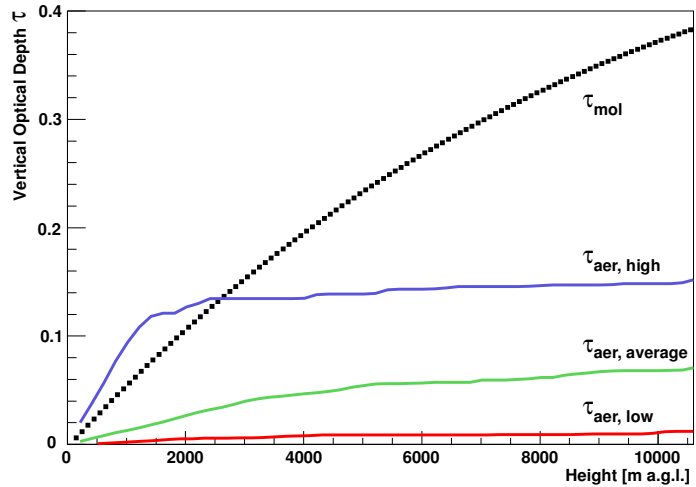}
    \caption{\label{fig:tauMolAer}
      The vertical profile of the molecular optical depth at 355~nm (dots),
      shown together with the measured vertical profiles of the aerosol optical
      depth in case of high, average, and low aerosol attenuation of the light.
      Height is measured above the ground.
    }
  \end{center}
\end{figure}

Similar to the aerosol transmission factor, the molecular transmission factor
for UV light at 355~nm can be calculated using the same geometry,
\begin{equation}
  T_{\rm mol}(h)
    = \exp{ \left[- ( \tau_{\rm mol}(h) / \sin{\varphi_2}) \right] }.
\end{equation}
In Fig.~\ref{fig:tauMolAer}, the vertical profile of the molecular optical depth
$\tau_{\rm mol}(h)$ is compared with measured aerosol profiles $\tau_{\rm
aer}(h)$ (Eq.~\ref{eqn:T_aer_alpha}) in case of high, average and low aerosols
attenuation of light in the air. We define ``high'' aerosol attenuation when
$\tau_{\rm aer}(5\rm km) > 0.1$, ``average'' when $0.04 < \tau_{\rm aer}(5\rm
km) < 0.05$ and ``low'' when $\tau_{\rm aer}(5\rm km) < 0.01$. Considering an
emission point $\rm P1$ at an altitude of 5 km and a distance on ground of 30 km
from the FD, the quoted high, average and low values correspond to transmission
factors of $T_{\rm aer} < 0.54$,  $0.73 < T_{\rm aer} < 0.78$ and $T_{\rm aer} >
0.94$, respectively.  The steps seen in the $\tau_{\rm aer}$ profiles are due to
multiple aerosol layers at different altitudes. For the calculation of the
molecular optical depth profile, monthly averaged temperature, pressure, and
humidity profiles for the location of the Observatory were used. The
12~resulting $\tau_{\rm mol}$ profiles were averaged, the fluctuations
introduced by the varying atmospheric state variables throughout the year are
very small, comparable to the size of the points in Fig~\ref{fig:tauMolAer}. On
the other hand, the aerosol attenuation can vary between clear and hazy
conditions within a few days, making the constant monitoring of the aerosol
optical depth necessary.

%% file: clf.tex
\section{The Central Laser Facility}
\label{sec:instr}

\begin{figure}[!ht]
  \centering
  \begin{minipage}[t]{.43\textwidth}
    \centering
    \includegraphics*[width=.99\linewidth,clip,trim=0cm 0cm 3cm 0cm]{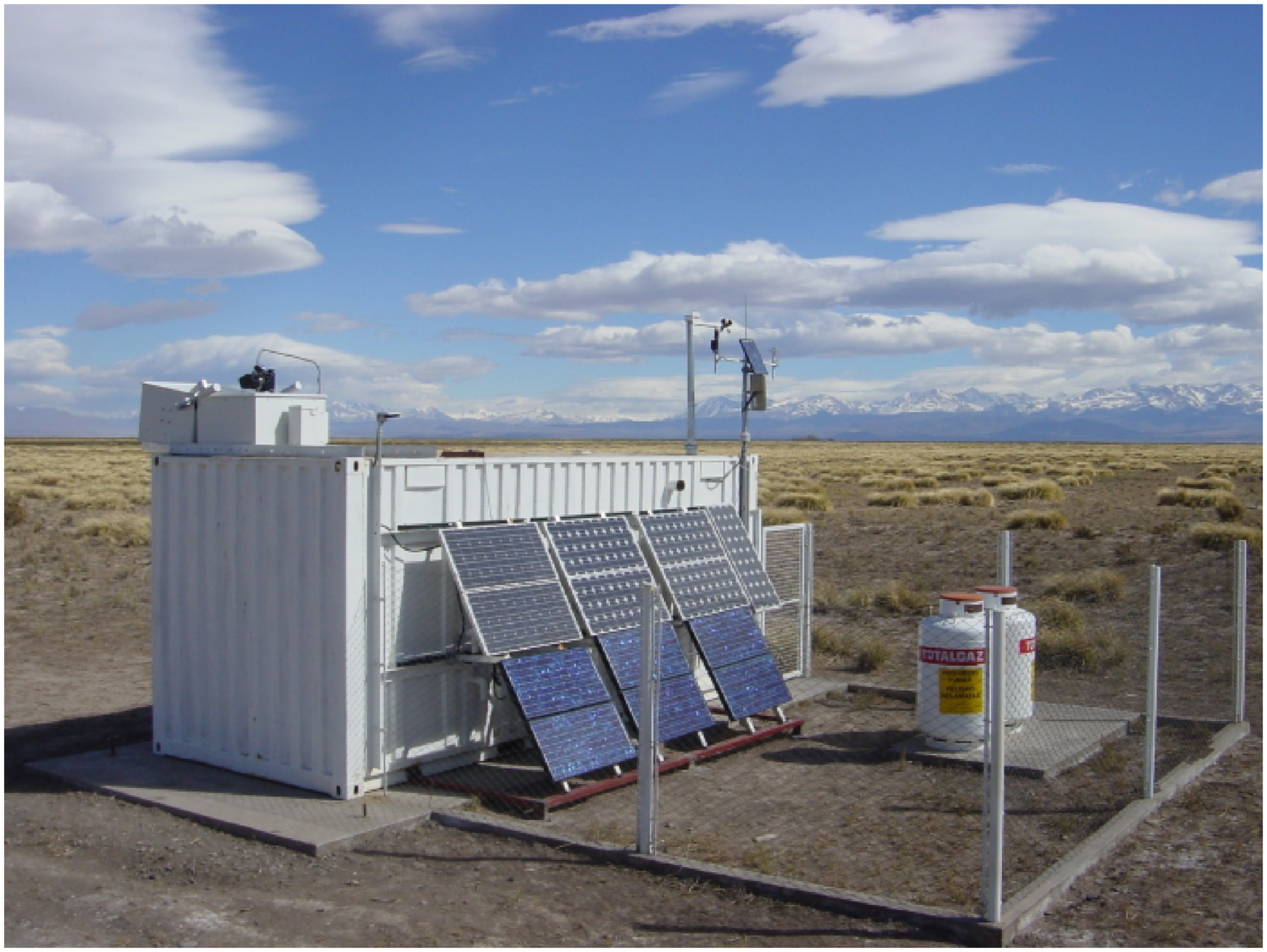}
  \end{minipage}
  \hfill
  \begin{minipage}[b]{.54\textwidth}
    \centering
    \includegraphics*[width=.99\linewidth,clip,trim=1cm 1.2cm 0cm 1.6cm]{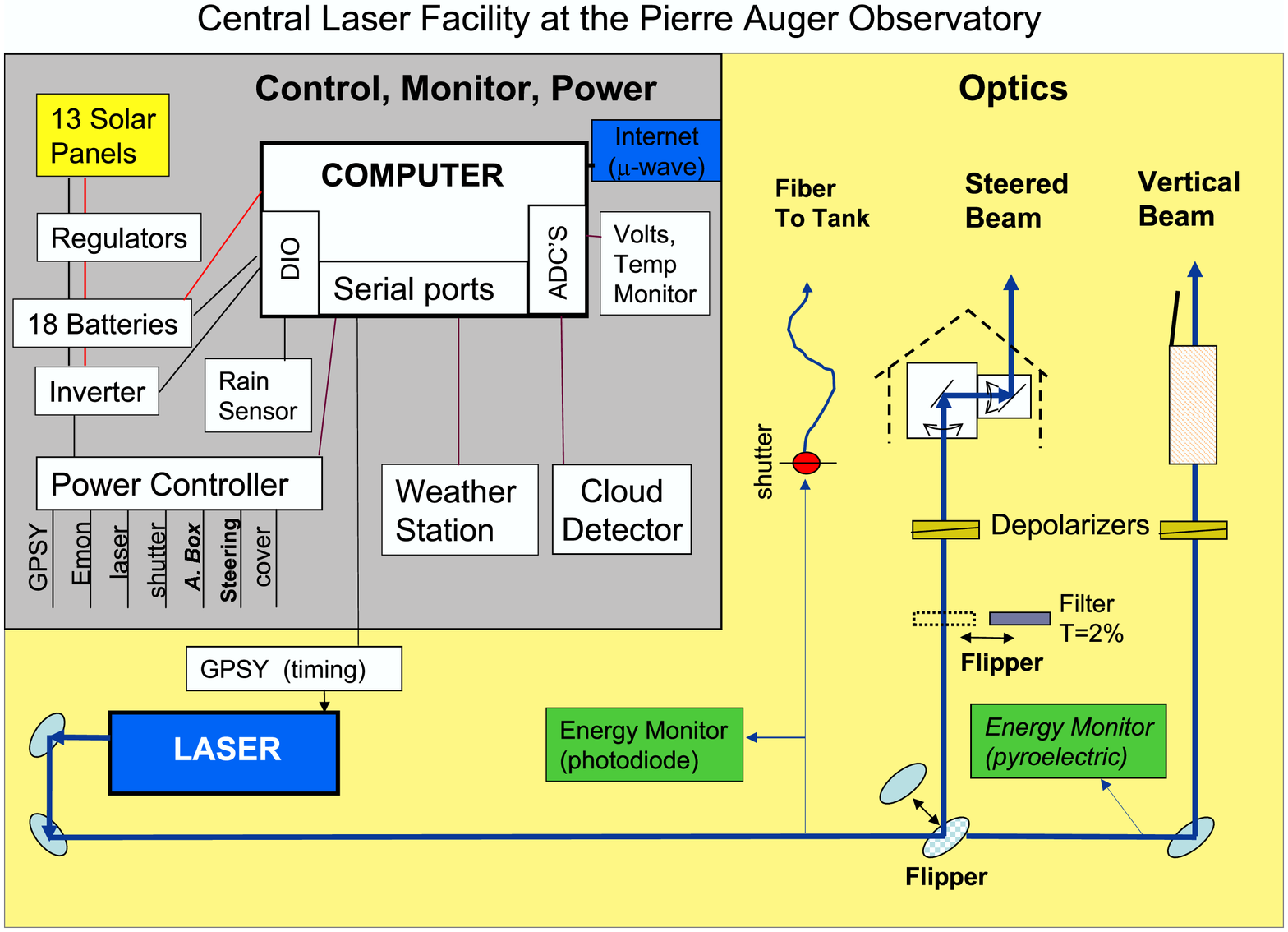}
  \end{minipage}
  \caption{\label{fig:clfphoto}
    Left: The Central Laser Facility.
    Right: A schematic of the Central Laser Facility.
  }
\end{figure}

The Central Laser Facility, described in detail elsewhere~\cite{clfjinst},
generates an atmospheric ``test beam''. Briefly, the CLF uses a frequency
tripled Nd:YAG laser, control hardware and optics to direct a calibrated pulsed
UV beam into the sky. Its wavelength of 355~nm is  near the center of the main
part of the nitrogen fluorescence spectrum~\cite{airfly}. The spectral purity of
the beam delivered to the sky is better than $99\%$. Light scattered from this
beam produces tracks in the FD telescopes. The CLF is located near the middle of
the array, nearly equidistant from three out of four of the FD sites, at an
altitude of 1416~m above sea level. The distances to the Los Leones (located
1416.2~m above sea level), Los Morados (1416.4~m), Loma Amarilla (1476.7~m) and
Coihueco (1712.3~m) FD sites are 26.0~km, 29.6~km, 40~km, and 30.3~km,
respectively. In Fig.~\ref{fig:clfphoto}, a picture (left) of the CLF is shown.
The CLF is solar-powered and operated remotely.

The laser is mounted on an optical table that also houses most of the other
optical components. The arrangement is shown in Fig.~\ref{fig:clfphoto} (right).
Two selectable beam configurations --~vertical and steerable~-- are available.
The steering mechanism consists of two mirrors on rotating, orthogonal axes
which can direct the beam in any direction above the horizon. The inclined laser
shots can be used to calibrate the pointing and time offsets of the fluorescence
telescopes. For the aerosol analyses described in this paper, only the vertical
beam is used. For this configuration, the beam direction is maintained within
0.04\degree of vertical with full-width beam divergence of less than
0.05\degree.

The Nd:YAG laser emits linearly polarized light. To perform the aerosol
measurements described in this paper, it is convenient, for reasons of symmetry,
to use a vertical beam that has no net polarization. In this case equal amounts
of light are scattered in the azimuthal directions of each FD site. Therefore,
the optical configuration includes depolarizing elements that randomize the
polarization by introducing a varying phase shift across the beam spot. The net
polarization of the fixed-direction vertical beam is maintained within 3$\%$ of
random.

The nominal energy per pulse is 6.5~mJ and the pulse width is 7~ns. Variations
in beam energy are tracked to an estimated accuracy of 3$\%$. The relative
energy of each vertical laser shot is independently measured by a photodiode and
a pyroelectric probe. The CLF laser energy is periodically calibrated and optics
are cleaned. For each of these periods a new coherent data set is defined and
the corresponding period referred to as a \emph{CLF epoch}. The length of an
epoch varies between a few months and one year.

The CLF fires 50~vertical shots at 0.5~Hz repetition rate every 15~minutes
during the FD data acquisition. Specific GPS timing is used to distinguish laser
from air shower events. The direction, time, and relative energy of each laser
pulse is recorded at the CLF and later matched to the corresponding laser event
in the FD data.

An upgrade~\cite{wiencke_upgrade} to the CLF is planned for the near future.
This upgrade will add a backscatter Raman LIDAR receiver, a robotic calibration
system, and replace the current flash lamp pumped laser by a diode pumped laser.

%% file: analysis.tex
\section{CLF Data Analysis}
\label{sec:analysis}

\begin{figure}[tb]
  \begin{center}
    \includegraphics*[width=0.6\textwidth,clip]{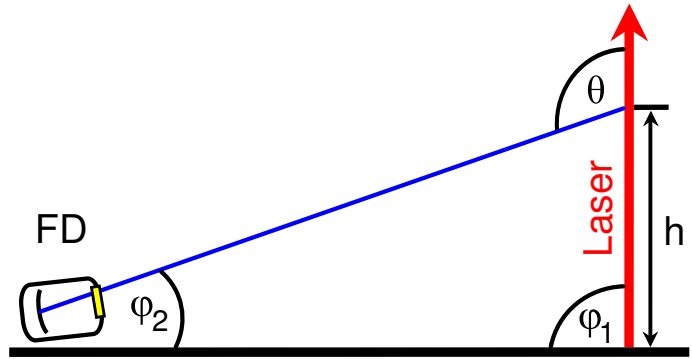}
    \caption{
      \label{fig:laser_geometry}
      Laser-FD geometry. The light is scattered out of the laser beam at a
      height~$h$ at an angle~$\theta$.
    }
  \end{center}
\end{figure}

\begin{figure}[htb]
  \begin{center}
    \includegraphics*[width=0.48\textwidth,clip]{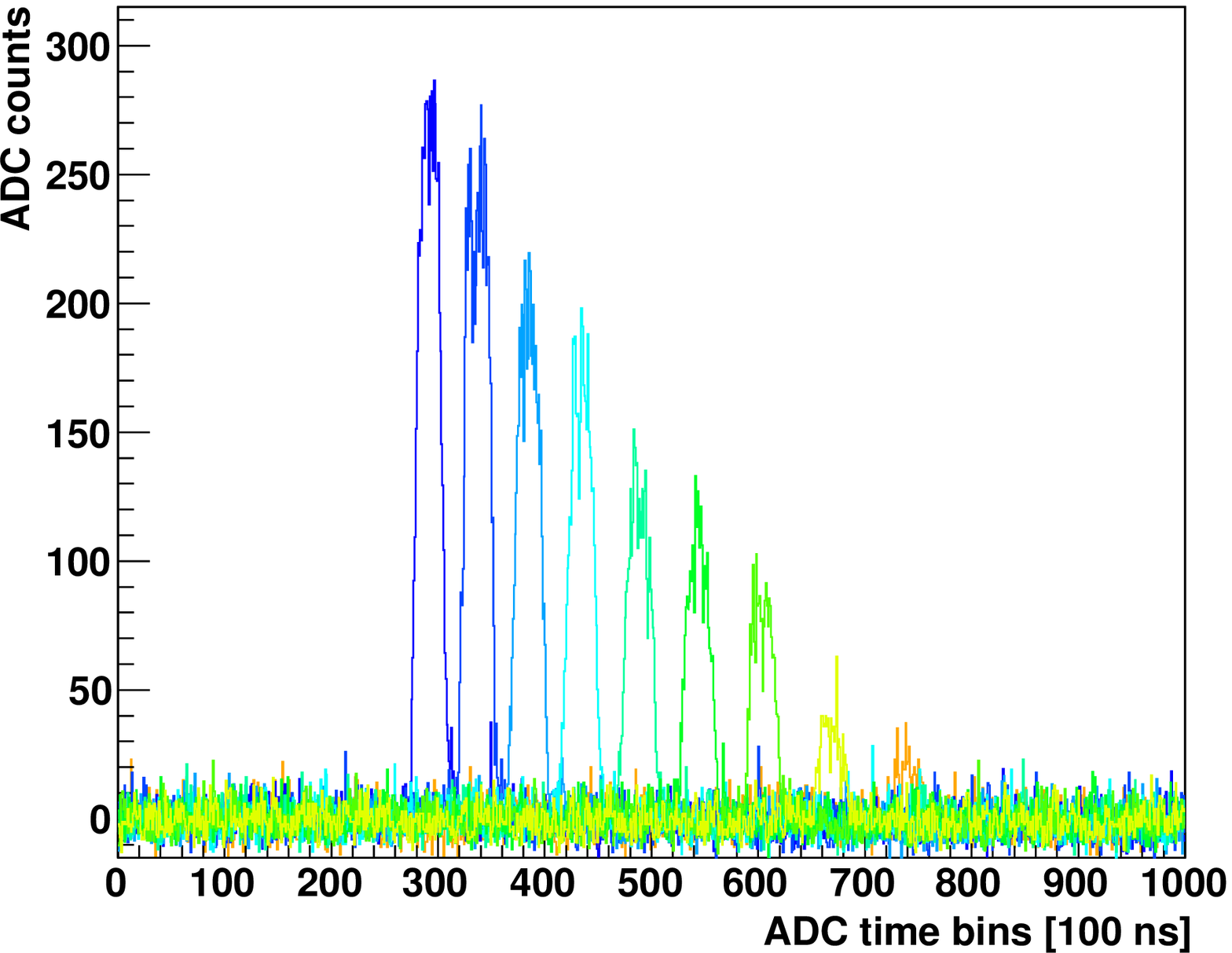}
    \includegraphics*[width=0.48\textwidth,clip]{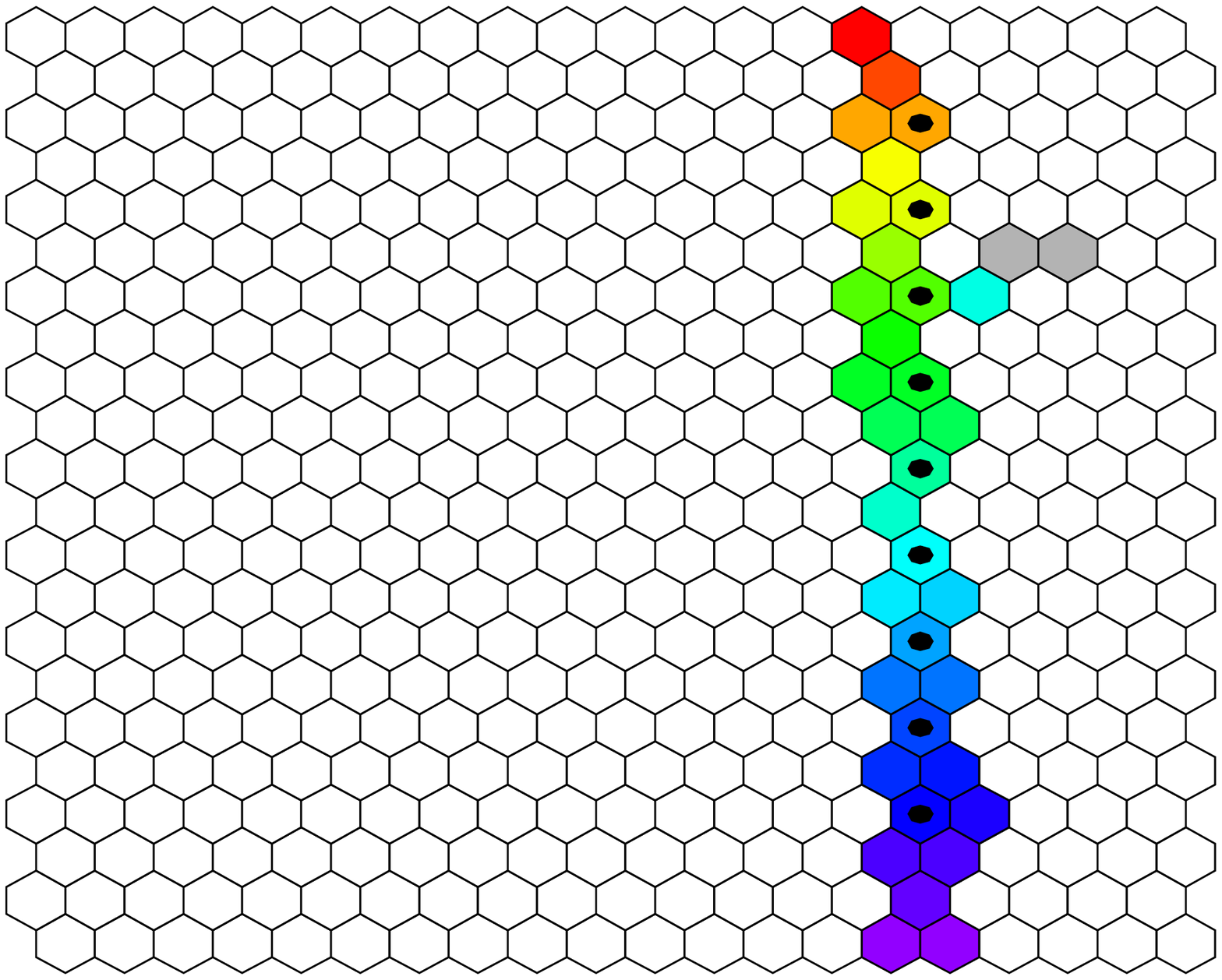}
    \caption{
      \label{fig:clf_fdeyedisplay}
      A 7~mJ CLF vertical event as recorded by the Los Leones FD site (distance
      26~km). Left panel: ADC counts vs.\ time (100~ns bins). The displayed data
      are for the marked pixels in the right panel. Right panel: Camera trace.
      The color code indicates the sequence in which the pixels were triggered.
    }
    \includegraphics*[width=0.48\textwidth,clip]{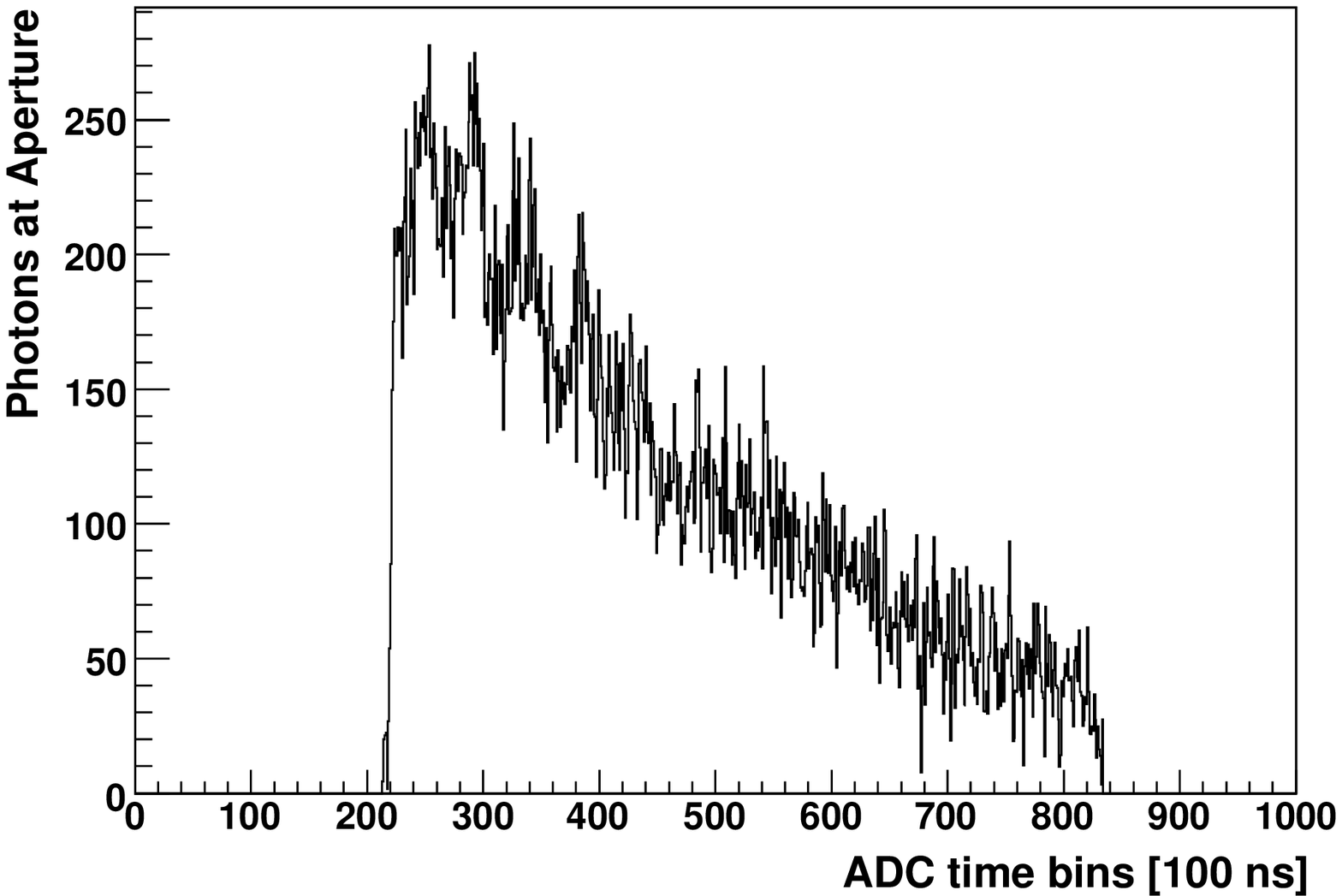}
    \includegraphics*[width=0.48\textwidth,clip]{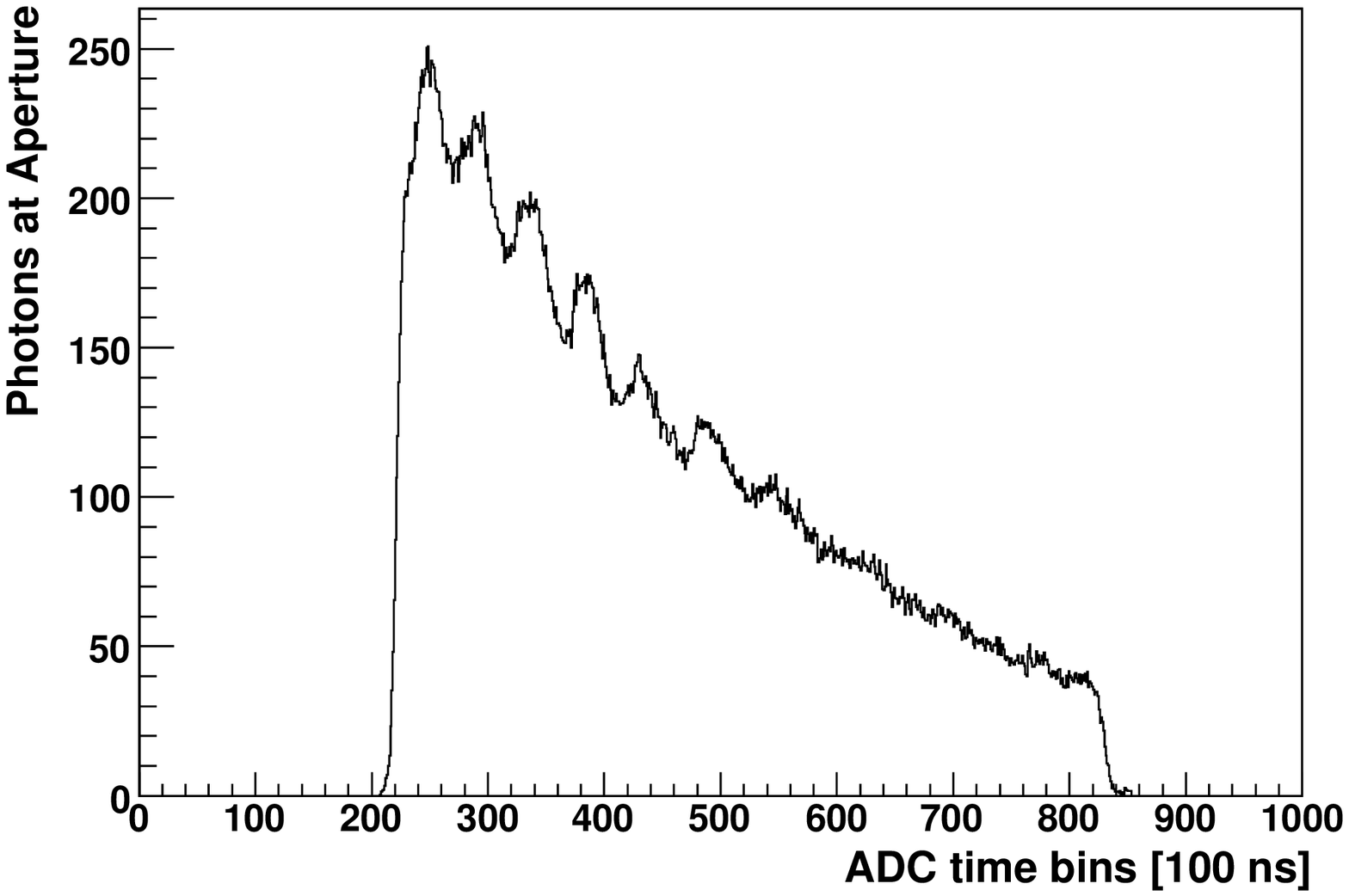}
    \caption[]{
      \label{fig:single_average_profile}
      Left: The light flux profile of a single CLF vertical shot seen from the
      Los Leones FD site. The same event as shown in
      Fig.~\ref{fig:clf_fdeyedisplay} is used. Right: 50~shots average profile.
    }
  \end{center}
\end{figure}

The light scattered out of the CLF laser beam is recorded by the FD (see
Fig.~\ref{fig:laser_geometry} for the laser-FD geometry layout). The angles from
the beam to the FD for vertical shots are in the range of 90\degree to
120\degree. As the differential scattering cross section of aerosol scattering
is much smaller than the Rayleigh scattering cross section in this range, the
scattering of light is dominated by well-known molecular processes. Laser tracks
are recorded by the telescopes in the same format used for air shower
measurements. In Fig.~\ref{fig:clf_fdeyedisplay}, a single 7~mJ CLF vertical
shot as recorded from the Los Leones FD site is shown. In the left panel of
Fig.~\ref{fig:single_average_profile}, the corresponding light flux profile for
the same event is shown. In Fig.~\ref{fig:single_average_profile}, right panel,
an average profile of 50~shots is shown.

\begin{figure}[htb]
  \begin{center}
    \includegraphics*[width=0.8\textwidth,clip]{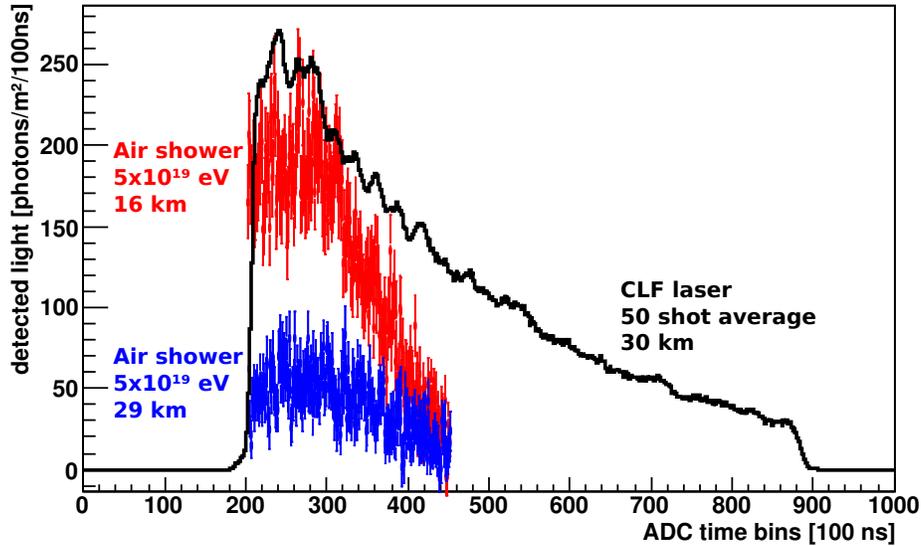}
    \caption{
      \label{fig:prof_laser_shower}
      Comparison between a 50 shot average of vertical 6.5~mJ UV laser shot from
      the CLF and near-vertical cosmic ray showers measured with the FD. The
      cosmic ray profile has been flipped in time so that in both cases the left
      edge of the profile corresponds to the bottom of the FD field of view.
    }
  \end{center}
\end{figure}

Laser light is attenuated in the same way as fluorescence light as it propagates
towards the FD. Therefore, the analysis of the amount of CLF light that reaches
the FD can be used to infer the attenuation due to aerosols. The amount of light
scattered out of a 6.5~mJ laser beam by the atmosphere is roughly equivalent to
the amount of UV fluorescence light produced by an EAS of $5 \times 10^{19}$~eV
at a distance to the telescope of about 16~km, as shown in
Fig.~\ref{fig:prof_laser_shower}. Also shown is the more attenuated light
profile of an almost identical shower at a larger distance.

Besides determining the optical properties of the atmosphere, the identification
of clouds is a fundamental task in the analysis of CLF laser shots. Clouds can
have a significant impact on shower reconstruction.

\begin{figure}[!t]
  \begin{minipage}[t]{.49\textwidth}
    \centering
    \includegraphics*[width=.99\linewidth,clip]{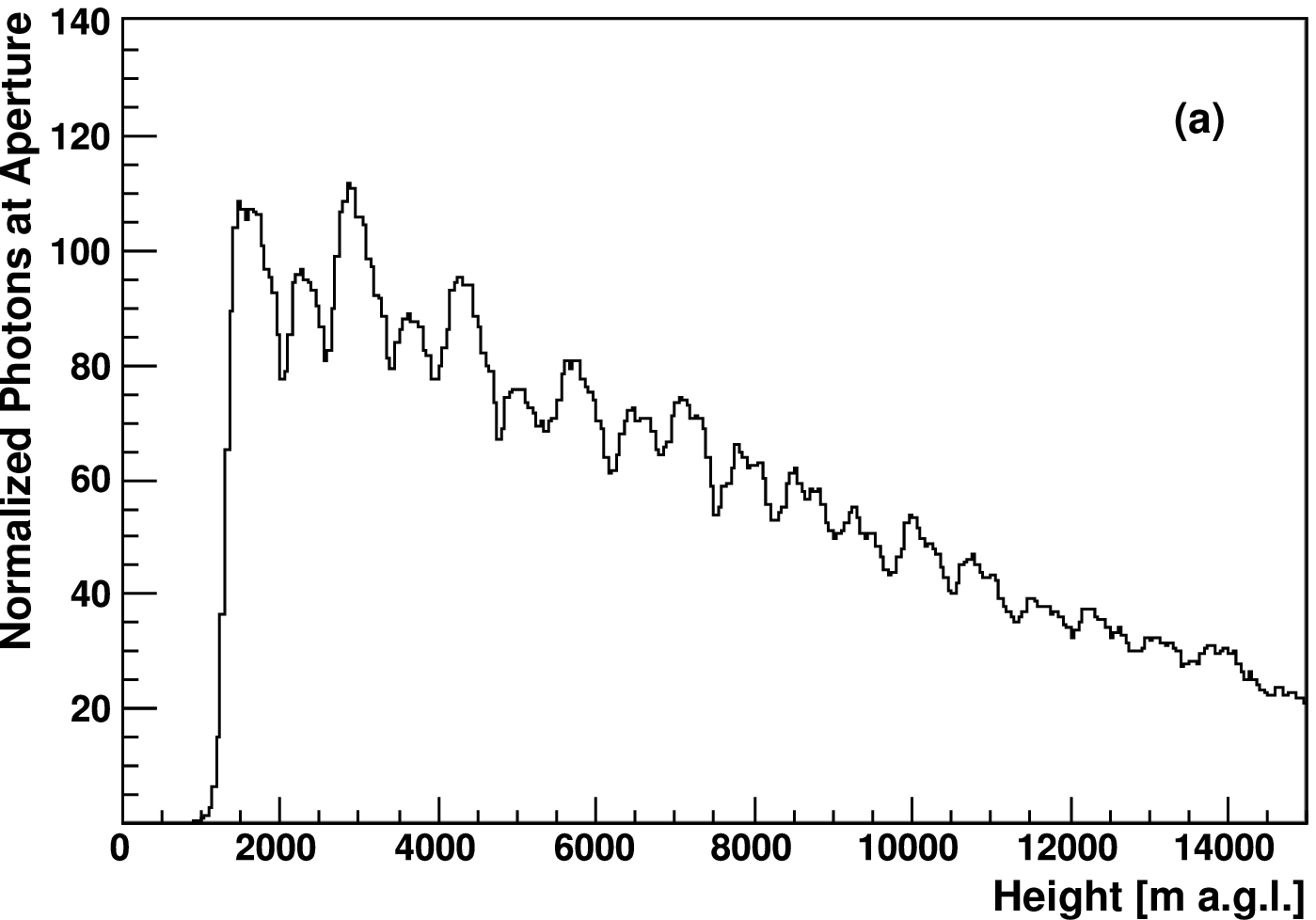}
  \end{minipage}
  \hfill
  \begin{minipage}[t]{.49\textwidth}
    \centering
    \includegraphics*[width=.99\linewidth,clip]{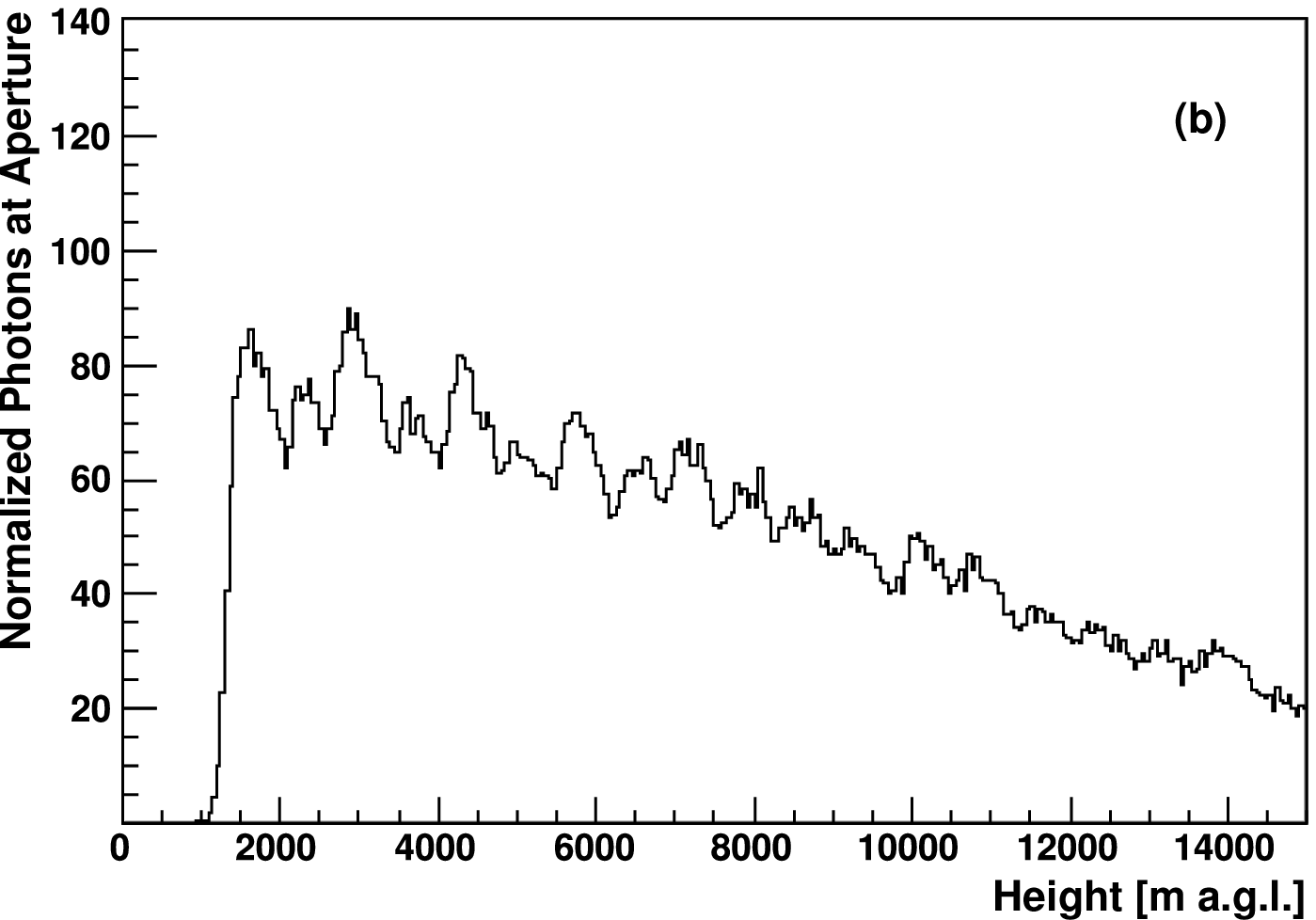}
  \end{minipage}
  \begin{minipage}[t]{.49\textwidth}
    \centering
    \includegraphics*[width=.99\linewidth,clip]{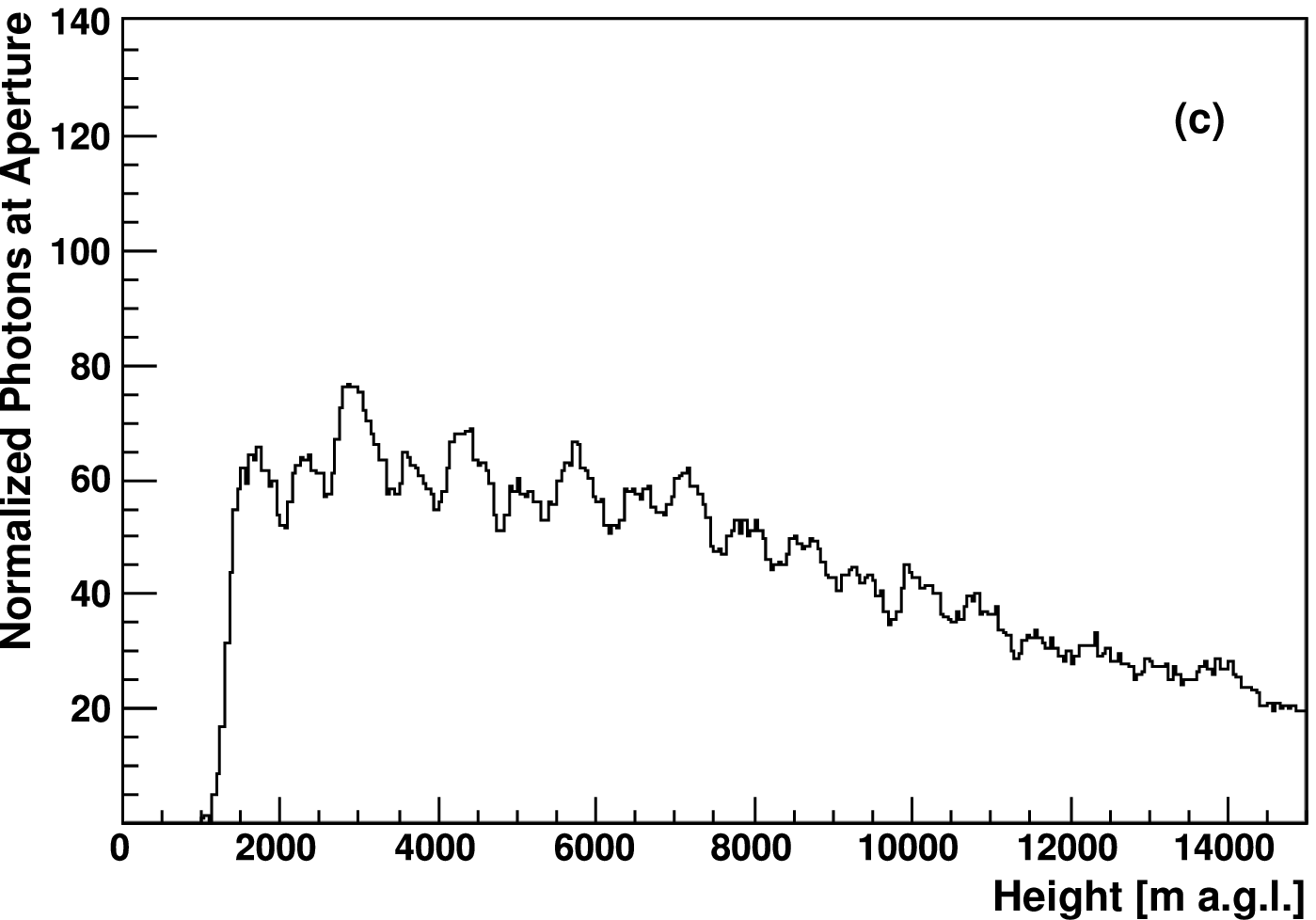}
  \end{minipage}
  \hfill
  \begin{minipage}[t]{.49\textwidth}
    \centering
    \includegraphics*[width=.99\linewidth,clip]{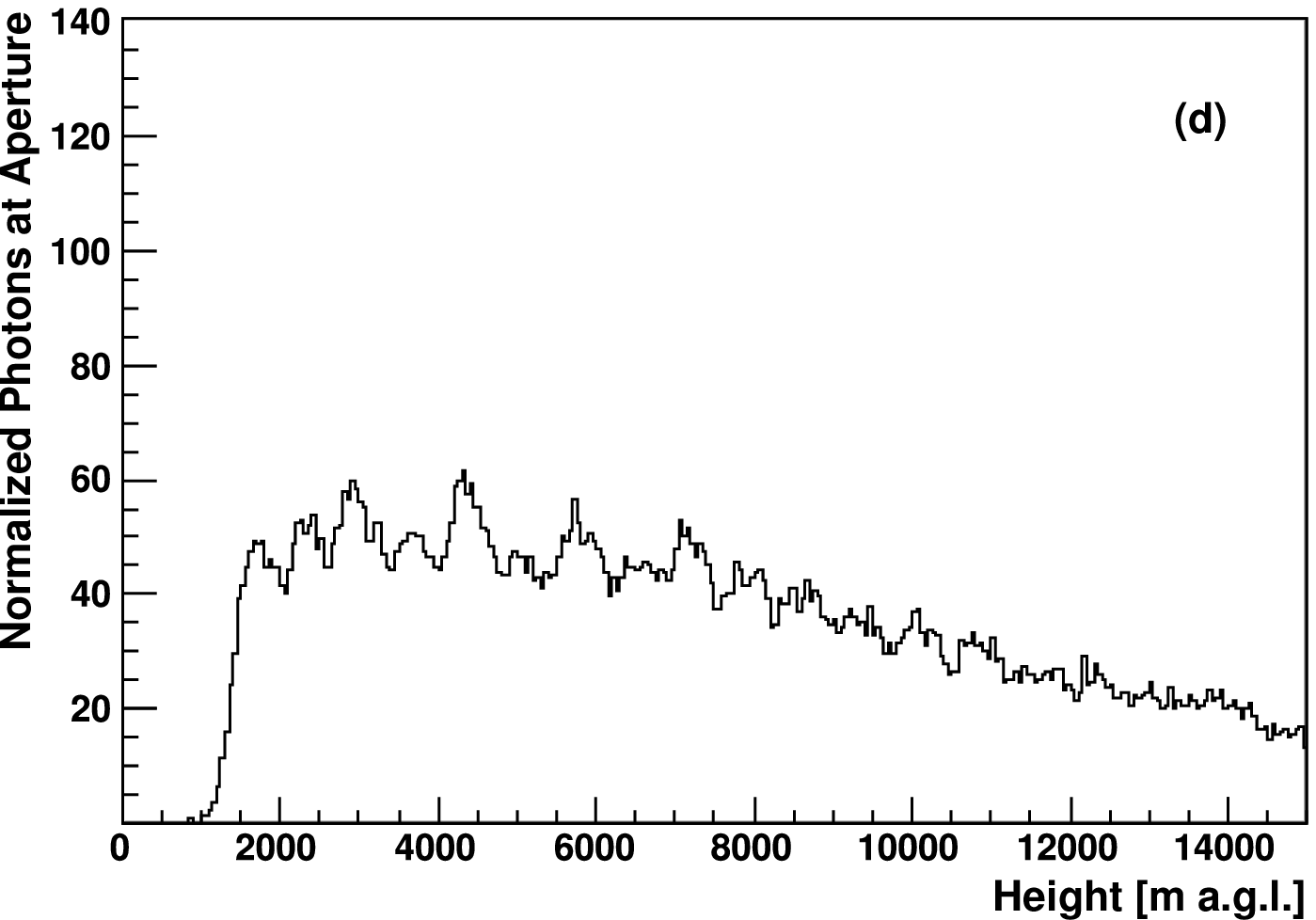}
  \end{minipage}
  \begin{minipage}[t]{.49\textwidth}
    \centering
    \includegraphics*[width=.99\linewidth,clip]{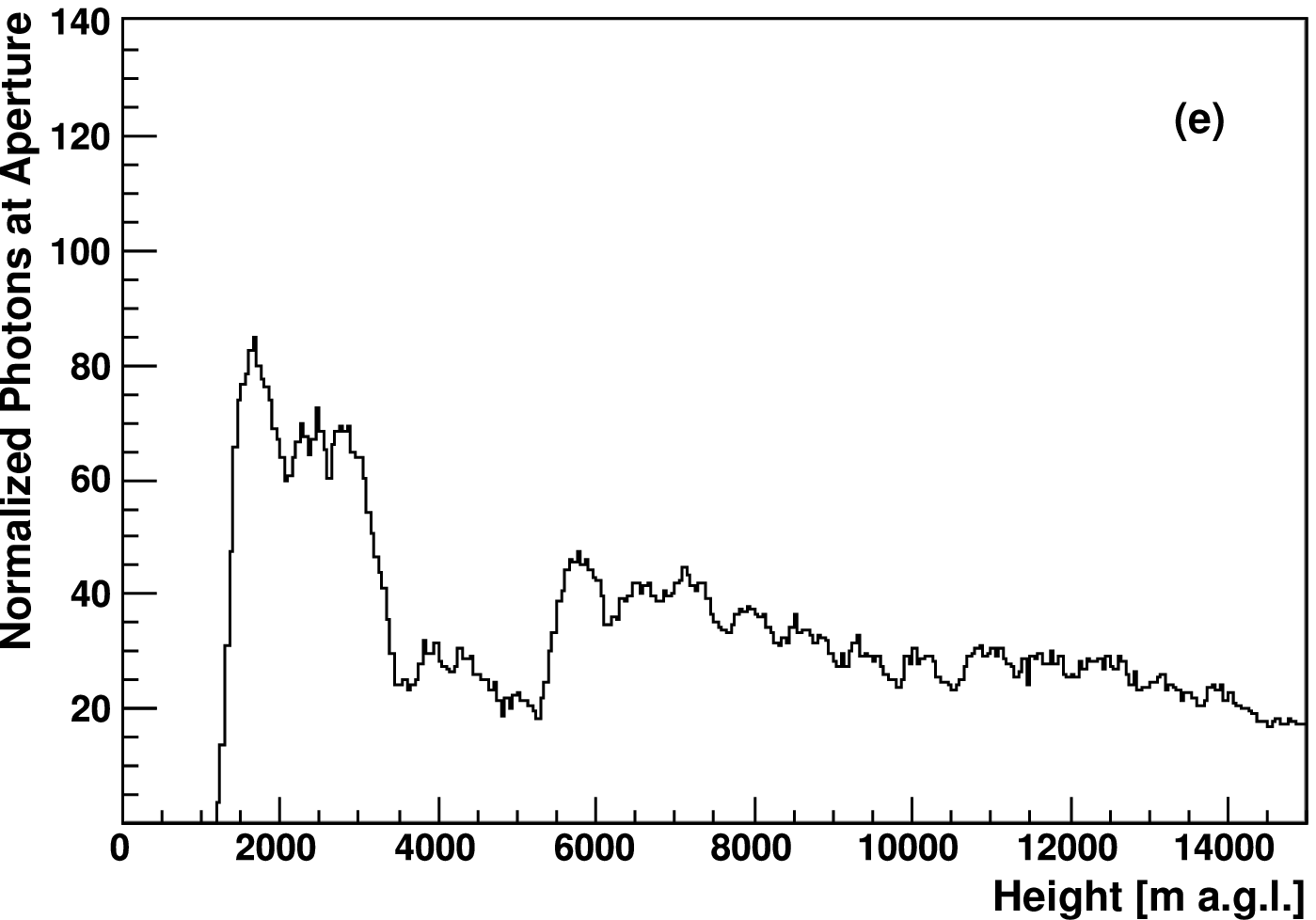}
  \end{minipage}
  \hfill
  \begin{minipage}[t]{.49\textwidth}
    \centering
    \includegraphics*[width=.99\linewidth,clip]{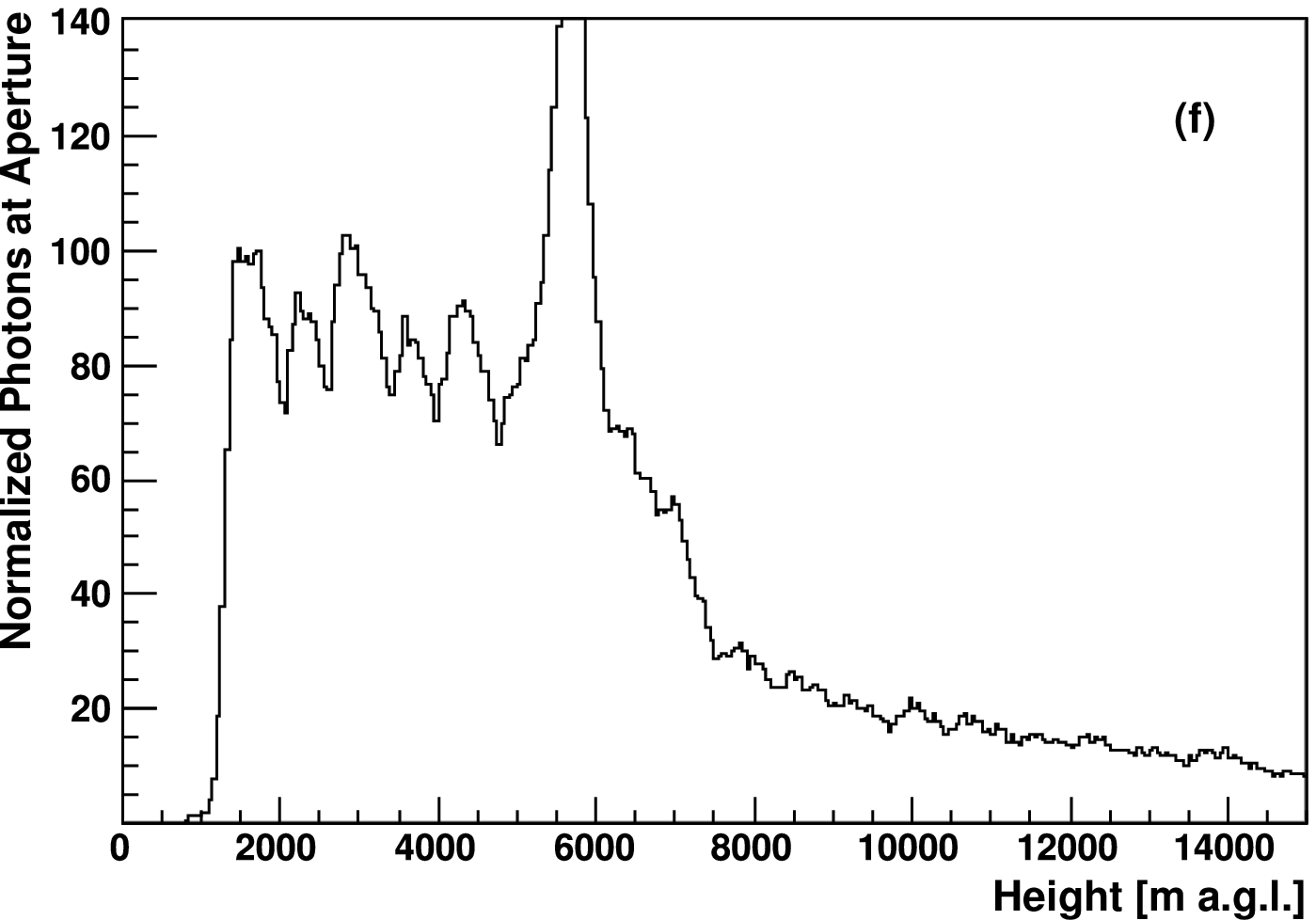}
  \end{minipage}
  \caption{\label{fig:HourlyExamples}
    Examples of light profiles measured with the FD at Coihueco under various
    atmospheric conditions. The height is given above the FD. The number of
    photons at the aperture of the FD is normalized per mJ of laser energy.
    Shown are a reference clear night (a); low (b), average (c) and high aerosol
    attenuation (d); cloud between FD and laser (e); laser beam passing through
    cloud (f).
  }
  \hfill
\end{figure}

In Fig.~\ref{fig:HourlyExamples}, examples of various hourly profiles affected
by different atmospheric conditions are shown. The modulation of the profile is
due to the FD camera structure, in which adjacent pixels are complemented by
light collectors. A profile measured on a night in which the aerosol attenuation
is negligible is shown in panel (a). Profiles measured on nights in which the
aerosol attenuation is low, average and high, are respectively shown in panels
(b), (c) and (d). As conditions become hazier, the integral photon count
decreases. The two bottom profiles (e) and (f) represent cloudy conditions.
Clouds appear in CLF light profiles as peaks or holes depending on their
position. A cloud positioned between the CLF and the FD can block the
transmission of light in its travel from the emission point towards the
fluorescence telescopes, appearing as a hole in the profile (e). The cloud could
be positioned anywhere between the CLF and the FD site, therefore its altitude
cannot be determined unambiguously. A cloud directly above the CLF appears as a
peak in the profile, since multiple scattering in the cloud enhances the amount
of light scattered towards the FD (f). In this case, it is possible to directly
derive the altitude of the cloud from the peak in the photon profile since the
laser-detector geometry is known.

Two independent analyses have been developed to provide hourly aerosol
characterization in the FD field of view using CLF laser shots from the
fixed-direction vertical configuration. To minimize fluctuations, both analyses
make use of average light flux profiles normalized to a fixed reference laser
energy.

\begin{itemize}
  \item The \emph{Data Normalized Analysis} is based on the comparison of measured
  profiles with a reference clear night profile in which the light
  attenuation is dominated by molecular scattering.
  \item The \emph{Laser Simulation Analysis} is based on the comparison of
  measured light flux profiles to simulations generated in various atmospheres
  in which the aerosol attenuation is described by a parametric model.
\end{itemize}

Measured profiles are affected by unavoidable systematics related to the FD and
laser calibrations. Simulated profiles are also affected by systematics related
to the simulation procedure. Using measurements recorded on extremely clear
nights where molecular Rayleigh scattering dominates, CLF observations can be
properly normalized without the need for absolute photometric calibrations of
the FD or laser. We will refer to these nights as \emph{reference clear nights}.
At present multiple scattering effects are not included in the laser simulation code, 
however the aforementioned normalization includes this effect for Rayleigh scattering, 
allowing to take it into account in the Laser Simulation Analysis.

\subsection{Reference clear nights}
\label{sec:raylnight}

In \emph{reference clear nights}, the attenuation due to aerosols is minimal
compared to the uncertainty of total attenuation, the scattering is dominated by
the molecular part. In such a clear night, the measured light profiles are
larger than profiles affected by aerosol attenuation, indicating maximum photon
transmission. Those profiles have shapes that are compatible with a profile
simulated under atmospheric conditions in which only molecular scattering of the
light is used. Reference clear night profiles are found by comparing measured
profiles to simulated average profiles of 50~CLF shots in a purely molecular
atmosphere at an energy of 6.5~mJ. Using the \mal Monthly Models described in
section \ref{sec:att}, the procedure is repeated 12~times using the appropriate
atmospheric density profiles.

The method chosen for the comparison is the unnormalized Kolmogorov-Smirnov
test. This test returns a pseudo-probability\footnote{the 
Kolmogorov-Smirnov test calculates probabilities for histograms containing 
counts, therefore here the returned value is defined as a pseudo-probability.} 
$P_{\rm KS}$ that the analyzed profile
is compatible with the clear one on the basis of shape only, without taking into
account the normalization. For each profile, $P_{\rm KS}$ and the ratio $R$
between the total number of photons of the measured profile and the simulated
clear one is calculated. In each CLF epoch, the search for the reference clear
night is performed among profiles having high values of $P_{\rm KS}$ and $R$. A
search region is defined by extracting the mean values $\mu_{P_{\rm KS}}$,
$\mu_R$ and the RMS $\sigma_{P_{\rm KS}}$,$\sigma_R$ of the distribution of each
parameter. Both parameters are required to be above their average $\mu +
\sigma$. Profiles belonging to the search region are grouped by night, and
nightly averages for the two parameters are computed $\left<P_{\rm KS}\right>$
and $\left<R\right>$. A list of candidate clear nights with associated
pseudo-probabilities and number of profiles is produced. The night with the highest
$\left<P_{\rm KS}\right>$ is selected and --~if available~-- at least
4~candidate profiles are averaged to smooth fluctuations. Once identified, the
associated $\left<R\right>$ is the normalization constant that fixes the energy
scale between real and simulated profiles needed in the Laser Simulation
Analysis. We estimated the uncertainty introduced by the method chosen to
identify the reference clear night by varying the cuts that determine the list
of candidate clear nights and the selection criteria that identify the chosen
reference night in the list. The normalization constant used to fix the energy
scale between real and simulated CLF profiles changes by less than 3$\%$.

As a final check to verify that the chosen nights are reference clear nights we
analyze the measurement of the aerosol phase function (APF)~\cite{apf} for that
night, measured by the APF monitor (see Sec.~\ref{sec:intro}). The molecular
part of the phase function $P_{\rm mol}(\theta)$ can be calculated analytically
from temperature, pressure and humidity at ground provided by weather stations.
After subtraction of the molecular phase function, the aerosol phase function
remains. In a reference clear night, the total phase function is dominated by
the molecular part with almost no contribution from aerosols. Since the APF
light source only fires approximately horizontally, this method to find the
reference nights is insensitive to clouds, so it can only be used as a
verification of reference nights that were found using the procedure described
in this section. After verification, the reference night is assumed to be valid
for the complete CLF epoch. In Fig.~\ref{fig:HourlyExamples}, panel (a), an
averaged light profile of a reference night is shown.

%% file: datanorm.tex
\subsection{Data Normalized Analysis}
\label{sec:DataNormalized}

\subsubsection{Building hourly laser profiles and cloud identification}

Using the timing of the event, the time bins of the FD data are converted to
height at the laser track using the known positions of the FD and CLF. The
difference in altitude between telescope and laser station and the curvature of
the Earth, which causes a height difference on the order of 50\,m, are taken
into account. The number of photons is scaled to the number of photons of a 1~mJ
laser beam (the normalization energy is an arbitrary choice that has no
implications on the measurements). The CLF fires sets of 50 vertical shots every
15 minutes. For each set, an average profile is built.

Clouds are then marked by comparing the photon transmission $T_{\rm aer}$ (see
Eq.~\ref{eqn:T_aer_alpha}) of the quarter hour profiles $T_{\rm quarter}$ to the
clear profile $T_{\rm clear}$ bin by bin. A ratio $T_{\rm quarter}/T_{\rm
clear}$ of less than 0.1 indicates a hole in the profile that is caused by a
cloud between the laser beam and the FD. A ratio larger than $1.3$ indicates
that the laser beam passed through a cloud directly above the CLF causing a
spike in the profile. In both cases, the minimum cloud height $h_{\rm cloud}$ is
set to the height corresponding to the lower edge of the anomaly. Only bins
corresponding to heights lower than this cloud height are used for the optical
depth analysis. Hours are marked as cloudy only if clouds are found in at least
two quarter hour sets, see Fig.~\ref{fig:ProfileExamples}. If there are no such
discontinuities, then $h_{\rm cloud}$ is set to the height corresponding to the
top of the FD camera field of view.

After $h_{\rm cloud}$ is determined, a preliminary full hour profile is made by
averaging all the available quarter hour profiles. One or more quarter hour
profiles can be missing due to the start or stop of FD data taking, heavy fog,
or problems at the CLF. Only one quarter hour profile is required to make a full
hour profile. Outlying pixels that triggered randomly during the laser event are
rejected and a new full hour profile is calculated. To eliminate outliers in
single bins that can cause problems in the optical depth analysis, the quarter
hour profiles are subjected to a smoothing procedure by comparing the current
profile to the preliminary full hour profile. After multiple iterations of this
procedure, the final full hour profile is constructed.

The maximum valid height $h_{\rm valid}$ of the profile is then determined. If
there is a hole in the profile of two bins or more due to the rejection of
outliers or clouds, $h_{\rm valid}$ is marked at that point. As with $h_{\rm
cloud}$, if no such holes exist, then $h_{\rm valid}$ is set to the height
corresponding to the top of the FD camera field of view. If $h_{\rm valid}$ is
lower than $h_{\rm cloud}$, the minimum cloud height is set to be the maximum
valid height. Points above $h_{\rm valid}$ are not usable for data analysis.

\subsubsection{Aerosol optical depth calculation}
\label{sec:TauCalcNorm}

Using the laser-FD viewing geometry shown in Fig.~\ref{fig:laser_geometry}, and
assuming that the atmosphere is horizontally uniform, it can be
shown~\cite{Abbasi} that the vertical aerosol optical depth is
\begin{equation}
\label{eq:geometric_vaod}
  \tau_{\rm aer}(h) =
    -\frac{\sin{\varphi_1}\sin{\varphi_2}}{\sin{\varphi_1}+\sin{\varphi_2}}
    \left(
      \ln\left(\frac{N_{\rm obs}(h)}{N_{\rm mol}(h)}\right) -
      \ln\left(1 + \frac{S_{\rm aer}(\theta, h)}{S_{\rm mol}(\theta,h)}\right)
    \right),
\end{equation}
where $N_{\rm mol}(h)$ is the number of photons from the reference clear profile
as a function of height, $N_{\rm obs}(h)$ is the number of photons from the
observed hourly profile as a function of height and $\theta$ is defined in
Fig.~\ref{fig:laser_geometry}. $S_{\rm aer}(\theta,h)$ and $S_{\rm
mol}(\theta,h)$ are the fraction of photons scattered out of the laser beam per
unit height by aerosols and air molecules, respectively. $S(\theta,h)$ is the
product of the differential cross section for scattering towards the FD
multiplied by the number density of scattering centers. For vertical laser shots
$(\varphi_1=\pi/2)$, $S_{\rm aer}(\theta,h)$ is small compared to $S_{\rm
mol}(\theta,h)$ because typical aerosols scatter predominately in the forward
direction. Thus the second term in Eq.~\ref{eq:geometric_vaod} can be neglected
to first order and Eq.~\ref{eq:geometric_vaod} becomes
\begin{equation}
\label{eq:clf_init_vaod}
  \tau_{\rm aer}(h) =
    \frac{\ln{N_{\rm mol}(h)} - \ln{N_{\rm obs}(h)}}{1 + \cosec{\varphi_2}}.
\end{equation}
With these simplifications, the CLF optical depth measurements depend only on
the elevation angle of each laser track segment and the number of photons from
the observed track and the reference clear profile. The aerosol optical depth
may be calculated directly from Eq.~\ref{eq:clf_init_vaod}.

$\tau_{\rm aer}$ is calculated for each bin in the hourly profile. The optical
depth at the altitude of the telescope is set to zero and is interpolated
linearly between the ground and the beginning of $\tau_{\rm aer}^{\rm meas}$
corresponding to the bottom of the field of view of the telescope. This
calculation provides a first guess of the measured optical depth $\tau_{\rm
aer}^{\rm meas}$, assuming that aerosol scattering from the beam does not
contribute to the track profile. While this is true for regions of the
atmosphere with low aerosol content, $\tau_{\rm aer}^{\rm meas}$ is only an
approximation of the true $\tau_{\rm aer}$ if aerosols are present. To overcome
this, $\tau_{\rm aer}^{\rm meas}$ is differentiated to obtain an estimate of the
aerosol extinction $\alpha_{\rm aer}(h)$ in an iterative procedure.

It is possible to find negative values of $\alpha_{\rm aer}$. They are most
likely due to statistical uncertainties in the fit procedure, or can be due to
systematic effects. As the laser is far from the FD site, the brightest measured
laser light profile, after accounting for relative calibrations of the FD and
the laser, occurs during a clear reference night. However, there are
uncertainties (see Sec.~\ref{sec:ErrorsNorm}) in the calibrations that track the
FD PMT gains and the CLF laser energy relative to the reference period.
Therefore, in some cases it is possible that parts of a laser light profile
recorded during a period of interest can slightly exceed the corresponding
profile recorded during a reference period. Typically, these artifacts occur
during relatively clear conditions when the aerosol concentration is low. The
effect could also happen if a localized scattering region, for example a small
cloud that was optically too thin to be tagged as a cloud, remained over the
laser and scattered more light out of the beam. However, since negative values
of $\alpha_{\rm aer}$ are unphysical, they are set to zero. Since the integrated
$\alpha_{\rm aer}$ values are renormalized to the measured $\tau_{\rm aer}^{\rm
meas}$ profile, this procedure does not bias the aerosol profile towards larger
values. The remaining values of $\alpha_{\rm aer}$ are numerically integrated to
get the fit optical depth $\tau_{\rm aer}^{\rm fit}$.
The final values for $\alpha_{\rm aer}$ and $\tau_{\rm aer}^{\rm fit}$ can be
used for corrections in light transmission during air shower reconstruction.

\begin{figure}[!ht]
  \begin{minipage}[t]{.99\textwidth}
    \centering
    \includegraphics*[width=.99\linewidth,clip]{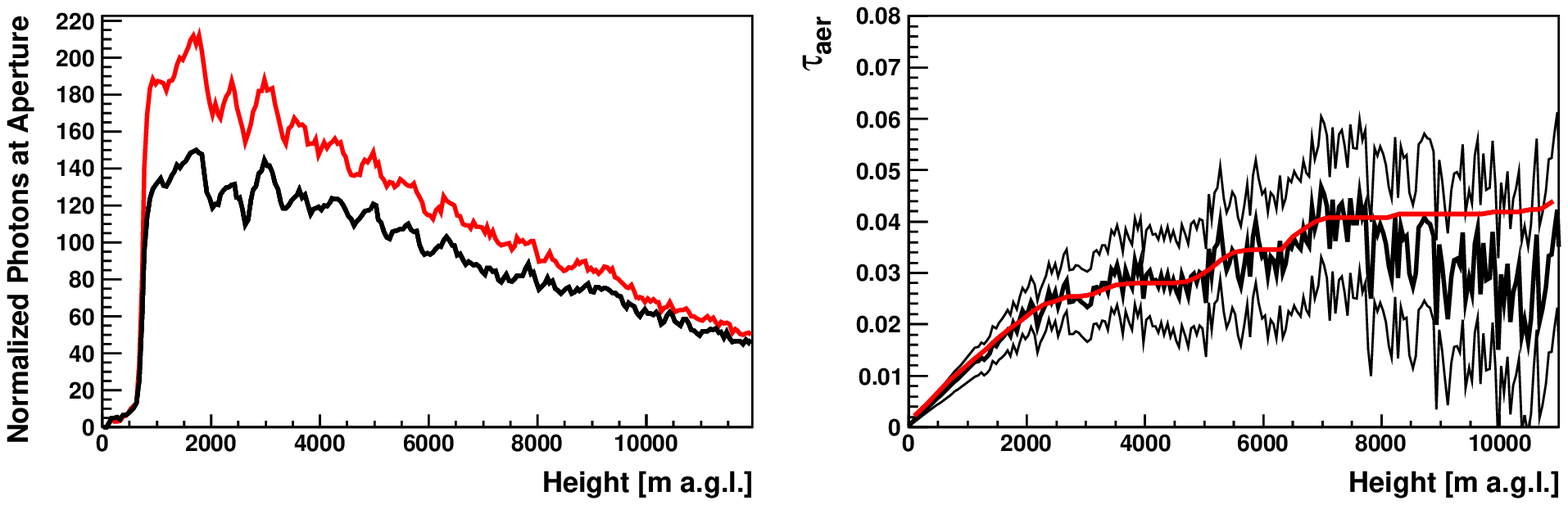}
  \end{minipage}
  \hfill
  \begin{minipage}[t]{.99\textwidth}
    \centering
    \includegraphics*[width=.99\linewidth,clip]{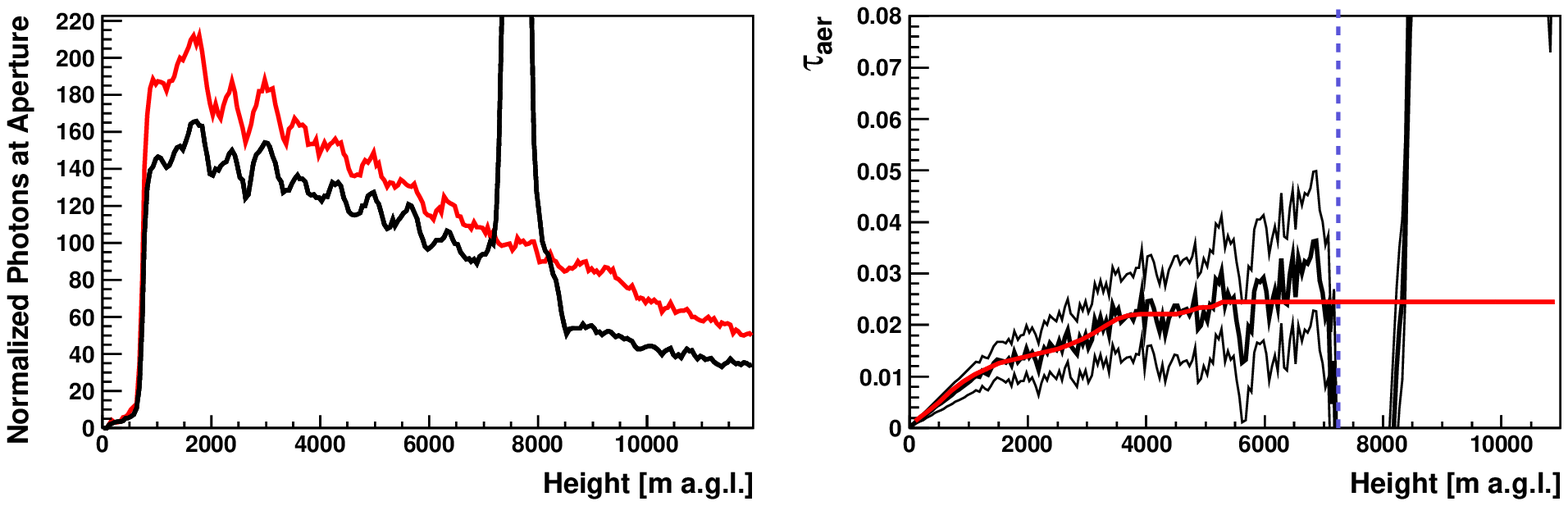}
  \end{minipage}
  \caption{\label{fig:ProfileExamples}
    Examples of light profiles and vertical aerosol optical depth $\tau_{\rm
    aer}$ measured with the FD at Los Morados during an average night (top) and
    with the laser passing through a cloud (bottom). The height is given above
    the FD, the light profile was normalized to a laser shot of 1~mJ. The black
    traces in left panels represent the hourly profiles, the red traces the
    reference clear nights. In the right panels, the thick black line represents
    $\tau_{\rm aer}^{\rm meas}$, the red line $\tau_{\rm aer}^{\rm fit}$.  The
    upper and lower traces correspond to the uncertainties. In the bottom right
    panel, the estimated cloud height is indicated by the vertical blue dotted
    line.
  }
\end{figure}

In Fig.~\ref{fig:ProfileExamples}, examples of laser and $\tau_{\rm aer}$
profiles are displayed from an average night and from a cloudy night when the
laser pulse passed through a cloud. In the left panels the black traces
represent the hourly profiles and the red traces represent the reference clear
nights. In the right panels $\tau_{\rm aer}^{\rm meas}$ and $\tau_{\rm aer}^{\rm
fit}$ measurements as a function of height are shown. The black curve is
$\tau_{\rm aer}^{\rm meas}$ and $\tau_{\rm aer}^{\rm fit}$ is overlaid in red.
The upper and lower traces correspond to the uncertainties. In the cloudy night,
a large amount of light is scattered by a cloud starting from a height of
approximately 7000~m. In the bottom right panel, the minimum height at which a
cloud was detected is indicated by a vertical blue line.

\subsubsection{Determination of Uncertainties}
\label{sec:ErrorsNorm}

Systematic uncertainties are due to uncertainty in the relative calibration of
the FD ($\sigma_{\rm cal}$), the relative calibration of the laser ($\sigma_{\rm
las}$), and the relative uncertainty in determination of the reference clear
profile ($\sigma_{\rm ref}$). A conservative estimate for each of these is
$3\%$. These uncertainties are propagated in quadrature for both the hourly
profile ($\sigma_{\rm syst,hour}$) and the clear profile ($\sigma_{\rm
syst,clear}$). The systematic uncertainty strongly depends on the height. Thus,
the viewing angle from the FD to the laser must be taken into account. The final
systematic uncertainty on $\tau_{\rm aer}^{\rm meas}$ is calculated by adding
$\sigma_{\rm syst,hour}$ and $\sigma_{\rm syst,clear}$ in quadrature, along with
the height correction,
\begin{equation}
\label{eq:final_systematic}
  \sigma_{\rm syst} =
    \frac{1}{1+\csc{\varphi_2}}\sqrt{(\sigma_{\rm syst,hour})^2
    + (\sigma_{\rm syst,clear})^2}.
\end{equation}
Two separate profiles are then generated corresponding to the values of
$\tau_{\rm aer}^{\rm meas} \pm \sigma_{\rm syst}$, as shown on the right panels
of Fig.~\ref{fig:ProfileExamples}.

The statistical uncertainty $\sigma_{\rm stat}$ is due to fluctuations in the
quarter hour profiles and is considered by dividing the RMS by the mean of all
quarter hour profiles at each height. These statistical uncertainties are
assigned to each bin of the $\tau_{\rm aer}^{\rm meas} \pm \sigma_{\rm syst}$
profiles. These two profiles are then processed through the same slope fit
procedure and integration as $\tau_{\rm aer}^{\rm meas}$ (see
Sec.~\ref{sec:TauCalcNorm}) to obtain the final upper and lower bounds on
$\tau_{\rm aer}^{\rm fit}$.

%% file: lasersim.tex
\subsection{Laser Simulation Analysis}
\label{naples}

\subsubsection{Atmospheric Model Description}

The atmospheric aerosol model adopted in this analysis is based on the assumption
that the aerosol distribution in the atmosphere is horizontally uniform.  The
aerosol attenuation is described by two parameters, the \emph{aerosol horizontal
attenuation length} $L_{\rm aer}$ and the \emph{aerosol scale height} $H_{\rm
aer}$.  The former describes the light attenuation due to aerosols at ground
level, the latter accounts for its dependence on the height.  
With this parameterization, the expression of the aerosol
extinction $\alpha_{\rm aer}(h)$ and the vertical aerosol optical depth
$\tau_{\rm aer}(h)$ are given by
\begin{equation}
\label{eqn_alpha}
  \alpha_{\rm aer}(h) = \frac{1}{L_{\rm aer}}
    \left[ \exp{ \left( -\frac{h}{H_{\rm aer}} \right) } \right],
\end{equation}
\begin{equation}
\label{eqnvaod}
  \tau_{\rm aer}(h_2-h_1) = \int_{h_1}^{h_2}{ \alpha_{\rm aer}(h) {\rm d} h } =
    - \frac{H_{\rm aer}}{L_{\rm aer}}
    \left[ \exp{ \left( -\frac{h_2}{H_{\rm aer} }\right) }
    - \exp{ \left( - \frac{h_1}{H_{\rm aer}} \right) }
    \right].
\end{equation}

Using Eq.~\ref{eqn:T_aer_alpha}, the aerosol transmission factor along the path
$s$ can be written as
\begin{equation}
  T_{\rm aer}(s) = \exp{ \left( \frac{H_{\rm aer}}{L_{\rm aer} \sin{\varphi_2}}
    \left[ \exp{ \left( - \frac{h_2}{H_{\rm aer}} \right) }
    - \exp{ \left( - \frac{h_1}{H_{\rm aer}} \right) } \right] \right) },
\end{equation}
where $h_1$ and $h_2$ are the altitudes above sea level of the first and second
observation levels and $\varphi_2$ is the elevation angle of the light path $s$
(cf.\ Fig.~\ref{fig:laser_geometry}).

The Planetary Boundary Layer (PBL) is the lower part of the atmosphere directly in contact with the ground, 
it is variable in height and the aerosol attenuation of light can be assumed as constant.  
The PBL is neglected in this two parameters approach. In the near future, the \emph{mixing layer height} 
will be introduced as a third parameter to take into account the PBL. In the Data Normalized Analysis, $\tau_{\rm aer}(h)$ 
is calculated per height bin in the hourly profile, therefore this analysis is sensible to the PBL 
and takes it into account.

\subsubsection{Building quarter-hour CLF profiles and generating a grid of simulations}
\label{sec:SimMethodDesc}

\begin{figure}[!ht]
  \begin{center}
    \includegraphics*[width=0.48\textwidth,clip]{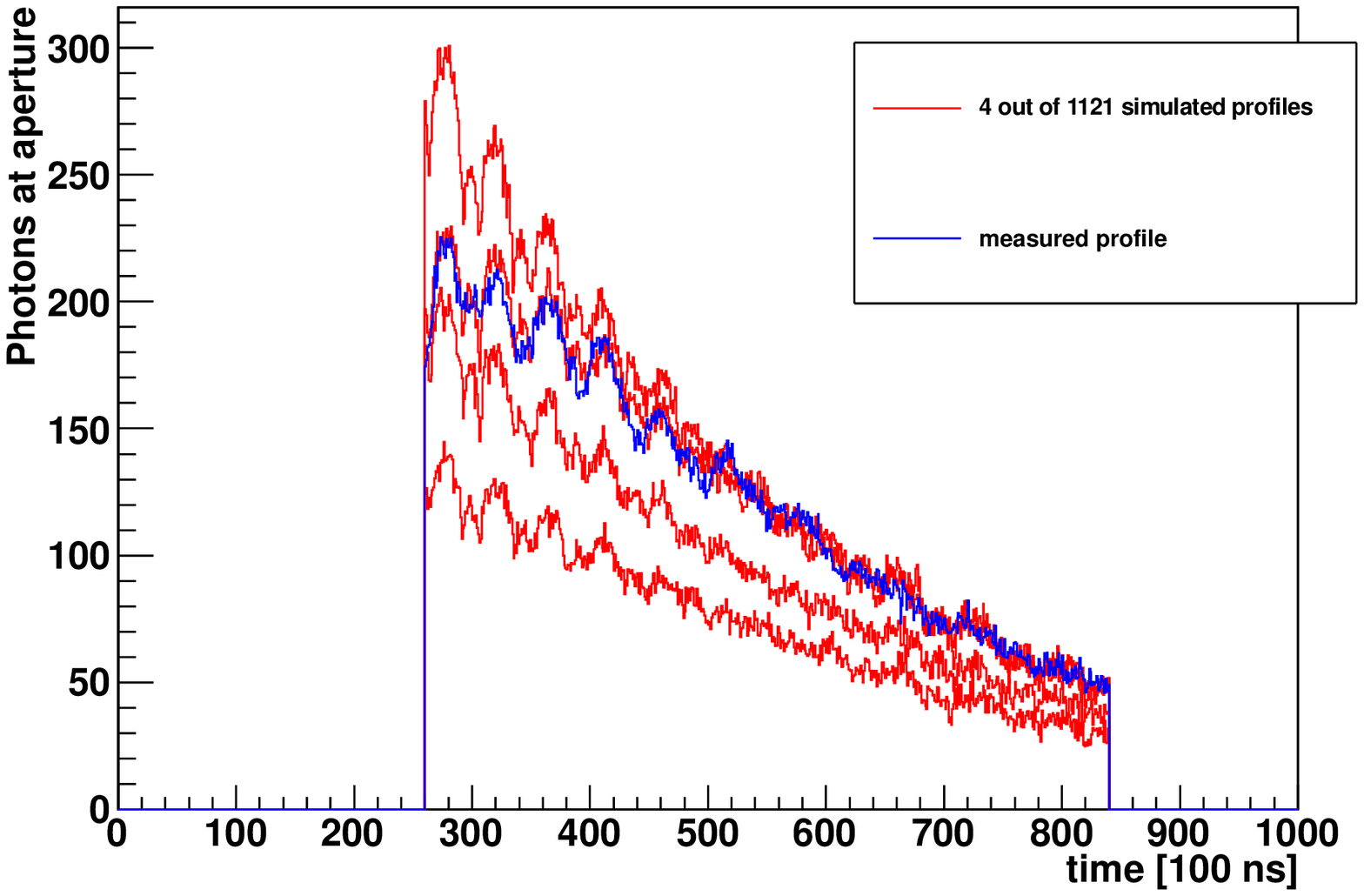}
    \includegraphics*[width=0.48\textwidth,clip]{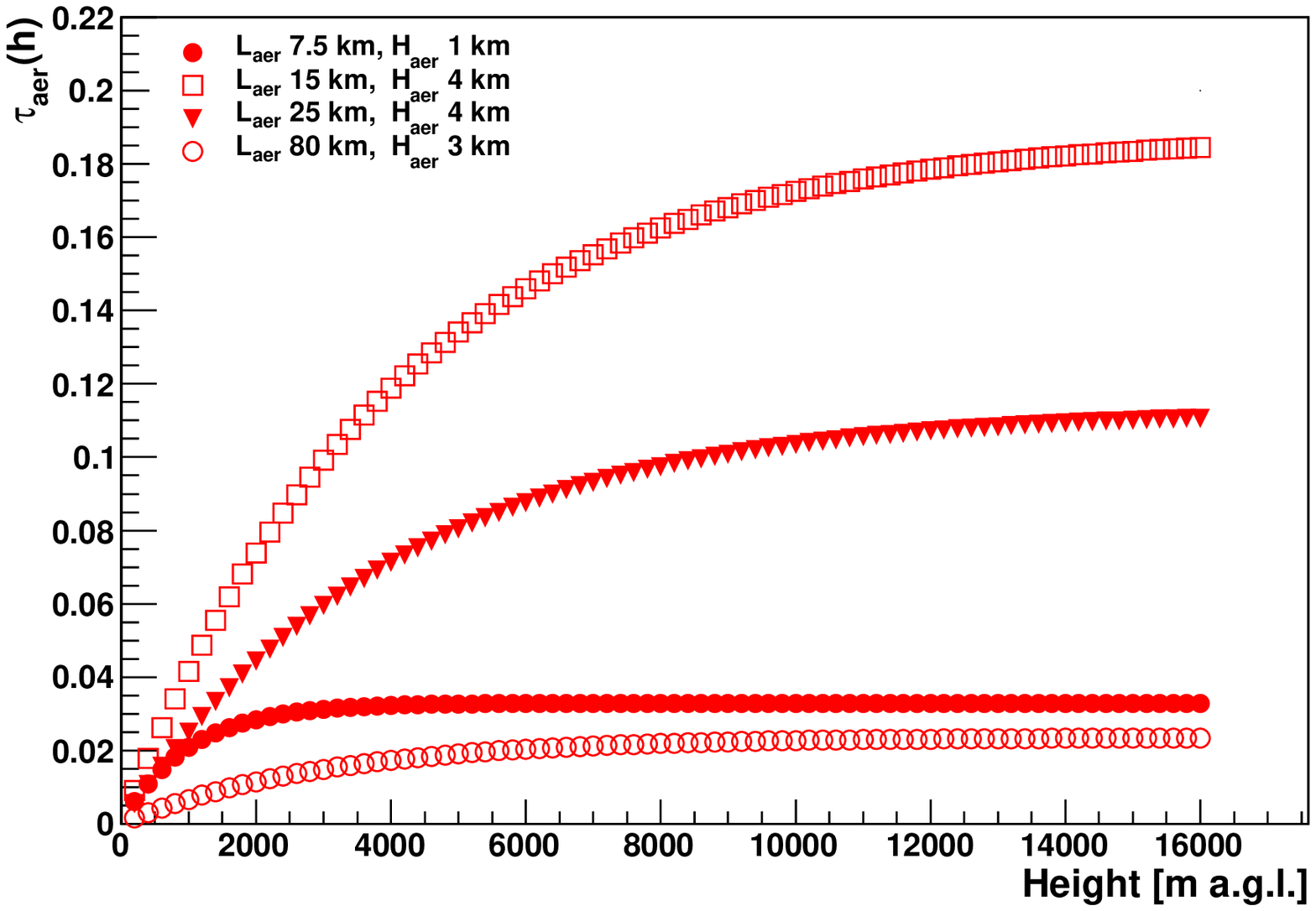}
    \caption{
      \label{fig:filmino_profili_vaod}
      Left: Four out of the 1\,121~simulated profiles of a monthly grid (red),
      superimposed to a measured profile (blue). Right: The four aerosol
      profiles corresponding to the simulated CLF profiles. In order, from top 
      to bottom, $\tau_{\rm aer}(h)$ profiles on the right correspond to CLF profiles 
      on the left from bottom to top.
    }
  \end{center}
\end{figure}

As described in section \ref{sec:instr}, the CLF fires 50 vertical shots every
15 minutes.  The profile of each individual event of the set is normalized to a
reference energy $E_{\rm ref}$, to compute an average profile equivalent to
$E_{\rm ref}$ for each group of 50 shots. In the following, this average light
profile will be referred to simply as ``profile''.  A grid of simulations at the
reference energy $E_{\rm ref}$ is generated, fixing the initial number of
photons emitted by the simulated vertical laser source.  While energy and
geometry of the simulated laser event are fixed, the atmospheric conditions,
defined by aerosol and air density profiles, are variable and described by means
of a two parameters models.  The aerosol attenuation profile in the atmosphere,
according to the model adopted, is determined setting values for $L_{\rm aer}$
and $H_{\rm aer}$.  For this analysis, the grid is generated by varying $L_{\rm
aer}$ from 5 to 150~km in steps of 2.5~km and $H_{\rm aer}$ from 0.5 km to 5~km
in steps of 0.25~km, corresponding to a total of 1121~profiles.  The air density
profiles are provided by the \mal Monthly Models, as discussed in
Sec~\ref{sec:att}.  Therefore, a total of 13\,452~profiles are simulated to
reproduce the wide range of possible atmospheric conditions on site. In the left
panel of Fig.~\ref{fig:filmino_profili_vaod}, a measured CLF profile (in blue)
is shown together with four out of the 1\,121~monthly CLF simulated profiles (in
red) used for the comparison procedure. In the right panel, the four aerosol
profiles~$\tau_{\rm aer}(h)$ corresponding to the simulated CLF profiles are
shown.

The relative energy scale between measured and simulated laser
profiles has to be fixed. The amplitude of CLF light profiles from laser shots
fired at the same energy depends on the aerosol attenuation in the atmosphere
and on absolute FD and CLF calibrations, that are known within 10$\%$ and 7$\%$,
respectively. The ratio of the amplitudes of the simulated clear night to the
measured reference clear night $R$ as defined in Sec.~\ref{sec:raylnight}
returns the normalization constant that fixes the relative energy scale between
measured and simulated laser profiles.  Using this normalization procedure, the
dependence on FD or CLF absolute calibrations is avoided and only the relative
uncertainty (daily fluctuations) of the laser probes (3$\%$) and FD calibration
constants (3$\%$) must be taken into account.  This procedure is repeated for
each CLF epoch data set. Average measured profiles are scaled by dividing the number of photons in 
each bin by the normalization constant of the corresponding epoch before measuring the aerosol attenuation.

\subsubsection{Optical depth determination and cloud identification}
\label{sec:OpticalDepthDet}

\begin{figure}[!ht]
  \begin{center}
    \includegraphics*[width=0.8\textwidth,clip]{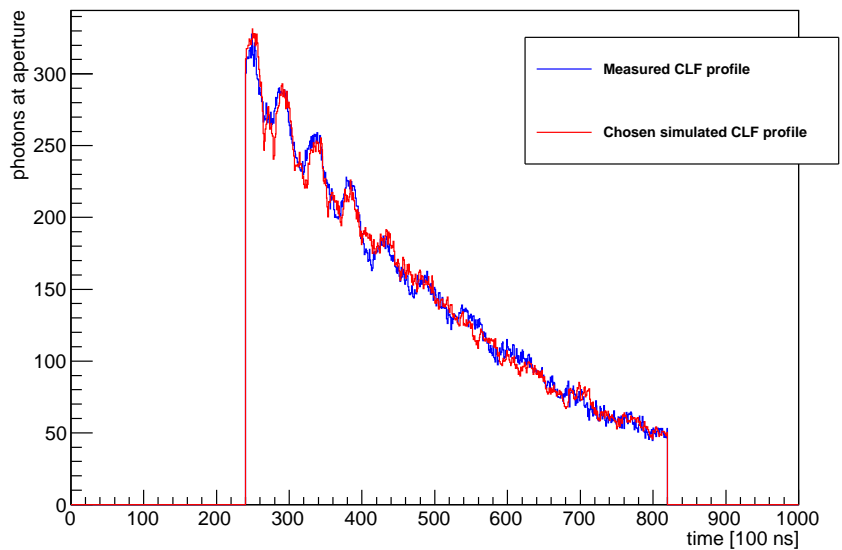}
    \caption{
      \label{fig:bestprofile}
      A measured CLF profile (blue) together with the chosen simulated (red).
	}
  \end{center}
\end{figure}

For each quarter hour average profile, the aerosol attenuation is determined
obtaining the pair $L_{\rm aer}^{\rm best}$, $H_{\rm aer}^{\rm best}$
corresponding to the profile in the simulated grid closest to the analyzed
event. The quantification of the difference between measured and simulated
profiles and the method to identify the closest simulation are the crucial 
points of this analysis. After validation tests on simulations of different 
methods, finally the pair $L_{\rm aer}^{\rm{best}}$ and 
$H_{\rm aer}^{\rm{best}}$ chosen is the one that
minimizes the square difference $D^2$ between measured and simulated profiles
computed for each bin, where $D^2 = [\sum_i(\Phi_i^{\rm meas} - \Phi_i^{\rm
sim})^2]$ and $\Phi_i$ are reconstructed photon numbers at the FD aperture in
each time bin.
In Fig.~\ref{fig:bestprofile}, an average measured profile as seen from Los
Leones compared to the simulated chosen profile is shown. The small discrepancy 
between measured and simulated profiles, corresponding to boundaries between 
pixels, has no effect on the measurements.

Before the aerosol optical depth is determined, the average profile is checked
for integrity and for clouds in the field of view in order to establish the
maximum altitude of the corresponding aerosol profile.  The procedure for the
identification of clouds works on the profile of the difference in photons for
each bin between the measured profile under study and the closest simulated
profile chosen from the grid. With this choice, the baseline is close to zero
and peaks or holes in the difference profile are clearly recognizable.  The
algorithm developed uses the bin with the highest or lowest signal and the
signal-to-noise ratio to establish the presence of a cloud and therefore
determines its altitude. The quarter hour information on the minimum cloud layer
height needed in the aerosol attenuation characterization is then stored.

If the average profile under study shows any anomaly or if a cloud is detected
between the laser track and the FD, it is rejected. If a cloud is detected above
the laser track, the profile is truncated at the cloud base height and this
lower part of the profile is reanalyzed, since the first search for clouds only
identifies the optically thicker cloud layer. If a lower layer of clouds is
detected in the truncated profile, or the cloud height is lower than 5500~m
a.s.l., the profile is rejected.

If no clouds are detected (either in the whole average profile or in the lower
part), the pair $L_{\rm aer}^{\rm best}$, $H_{\rm aer}^{\rm best}$, together
with the maximum height of the profile are stored and the procedure is
completed. The quarter hour $\tau_{\rm aer}(h)$ profile is calculated according
to Eq.~\ref{eqnvaod} together with the associated statistical and systematic
uncertainties. The information is stored, and the quarter hour $\tau_{\rm
aer}(h)$ profiles are averaged to obtain the hourly vertical aerosol optical
depth profile and the aerosol extinction profile $\alpha_{\rm aer}(h)$.

\subsubsection{Determination of Uncertainties}

Uncertainties on the vertical aerosol optical depth $\tau_{\rm aer}(h)$ are due
to the choice of the reference clear night, to the assumption that a parametric
model can be adopted to describe the aerosol attenuation, to the relative
uncertainty of nightly FD calibration constants --~converting ADC counts to
photon numbers~-- and CLF calibration constants --~converting laser probe
measurements to laser energy, and to the method used to choose the best matching
simulated profile.

To estimate the total uncertainty, the different contributions mentioned above
are evaluated and summed in quadrature. The uncertainty on the choice of the
reference clear night and the relative FD and CLF calibrations directly affect
the light profile, therefore they are summed in quadrature to estimate their
total contribution to the uncertainty on the photon profile, which is then
propagated to the aerosol profile. The uncertainty introduced by the method used
to identify the reference clear night is quoted at 3\% as described in
Sec.~\ref{sec:raylnight}; the contributions arising from the daily variations on
the FD and CLF calibration constants are both quoted at 3\%
level~\cite{clfjinst,FD}. Therefore, the total uncertainty of the number of
photons in the profile is less than 5.2\%. The effect on the aerosol profile
$\tau_{\rm aer}(h)$ of this total uncertainty on the light profile is evaluated
by increasing and decreasing the number of photons in the current CLF profile by
5.2\% and searching for the corresponding $\tau_{\rm min}(h)$ and $\tau_{\rm
max}(h)$ profiles. At each height, the error bars are given by $\tau_{\rm
best}(h) - \tau_{\rm min}(h)$ and $\tau_{\rm max}(h) - \tau_{\rm best}(h)$.

The contribution due to the parametric description of the aerosol attenuation of
light was determined comparing the hourly vertical aerosol optical depth
profiles obtained with the Laser Simulation Analysis to the corresponding
profiles obtained with the Data Normalized Analysis, which is not using a
parametric model for the aerosol attenuation. This comparison for each height
shows that aerosol profiles are compatible within 2\% at each altitude.

The uncertainty related to the method defined to choose the best matching
simulated profile as a function of the altitude is also estimated. As described
in Sec.~\ref{sec:OpticalDepthDet}, the parameters $L_{\rm aer}^{\rm{best}}$ and
$H_{\rm aer}^{\rm{best}}$ minimize the quantity $D^2 = [\sum_i(\Phi_i^{\rm real}
- \Phi_i^{\rm sim})^2]$. The method is repeated a second time in order to find
the couple $L_{\rm aer}^{\rm{err}}$ and $H_{\rm aer}^{\rm{err}}$ corresponding
to the quantity $D^{2\prime}$ nearest to $D^2$.  This profile is used to
estimate $\tau_{\rm err}(h)$, the uncertainty of the aerosol profile.
Therefore, the uncertainty related to the method $\sigma_{\rm method}(h)$
associated with $\tau_{\rm aer}(h)$ for each height bin is given by the
difference $\tau_{\rm best}(h) - \tau_{\rm err}(h)$. This uncertainty is
negligible with respect to the previous contributions.

The Laser Simulation Analysis extrapolates the aerosol attenuation for each
quarter hour CLF profile; then the four measured aerosol profiles are averaged
to obtain the hourly information needed for the air shower reconstruction. The
same procedure is adopted to obtain the uncertainties related to the hourly
aerosol attenuation profile. As a final step, the hourly uncertainty on
$\tau_{\rm aer}(h)$ is propagated to the aerosol extinction $\alpha_{\rm
aer}(h)$.

%% file: comparison.tex
\section{Comparison of the two analyses}
\label{sec:comparison}

The two analyses described in this paper independently produce hourly aerosol
profiles. In the Data Normalized Analysis, measured laser light profiles are
compared with an averaged light profile of a reference clear night. The Laser
Simulation Analysis is a procedure based on the comparison of CLF laser light
profiles with those obtained by a grid of simulated profiles in different
parameterized atmospheric conditions.

\begin{figure}[t]
  \begin{minipage}[t]{.49\textwidth}
    \centering
    \includegraphics*[width=0.99\textwidth,clip]{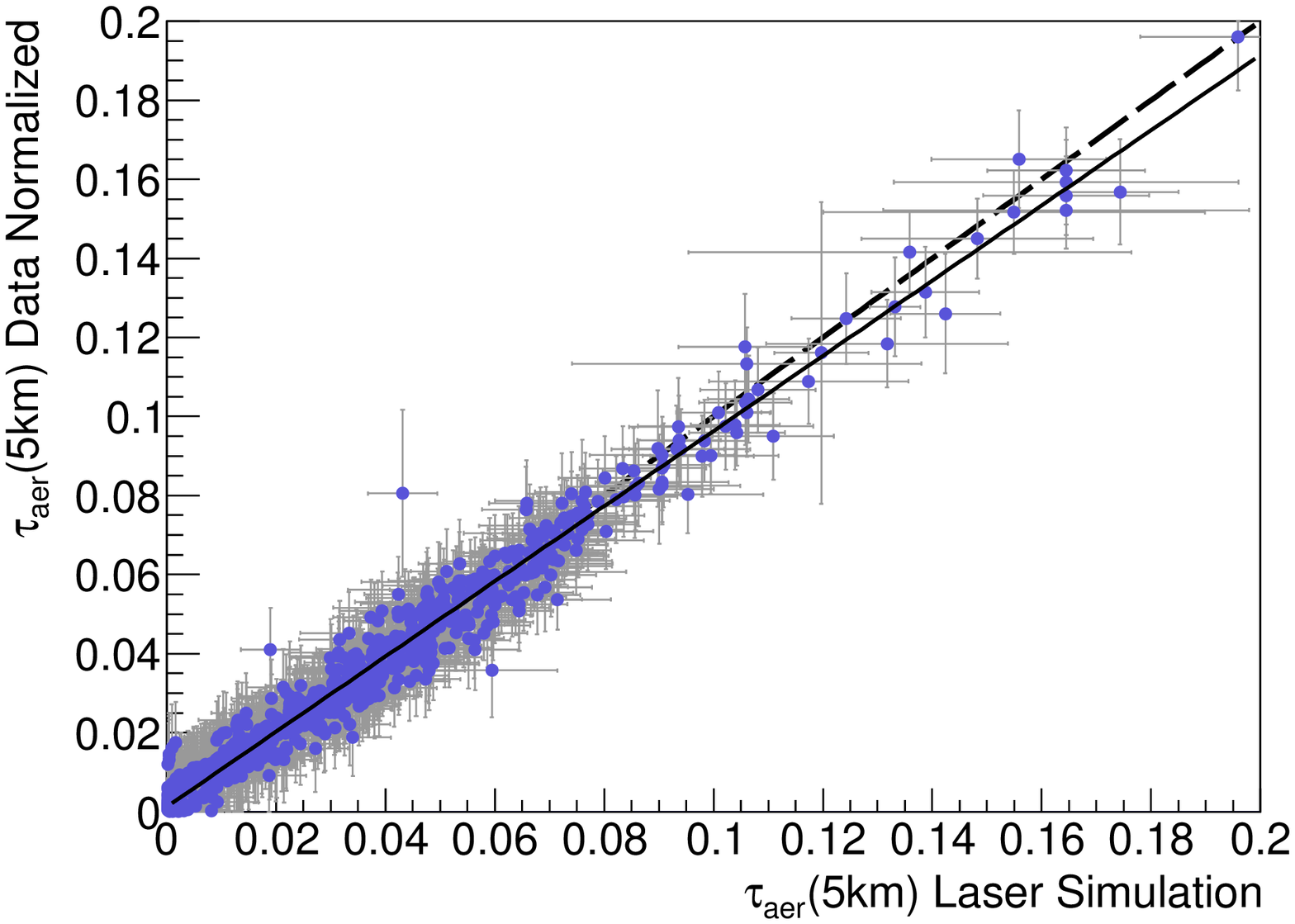}
    \caption*{(a) Correlation between the analyses.}
  \end{minipage}
  \hfill
  \begin{minipage}[t]{.49\textwidth}
    \centering
    \includegraphics*[width=0.99\textwidth,clip]{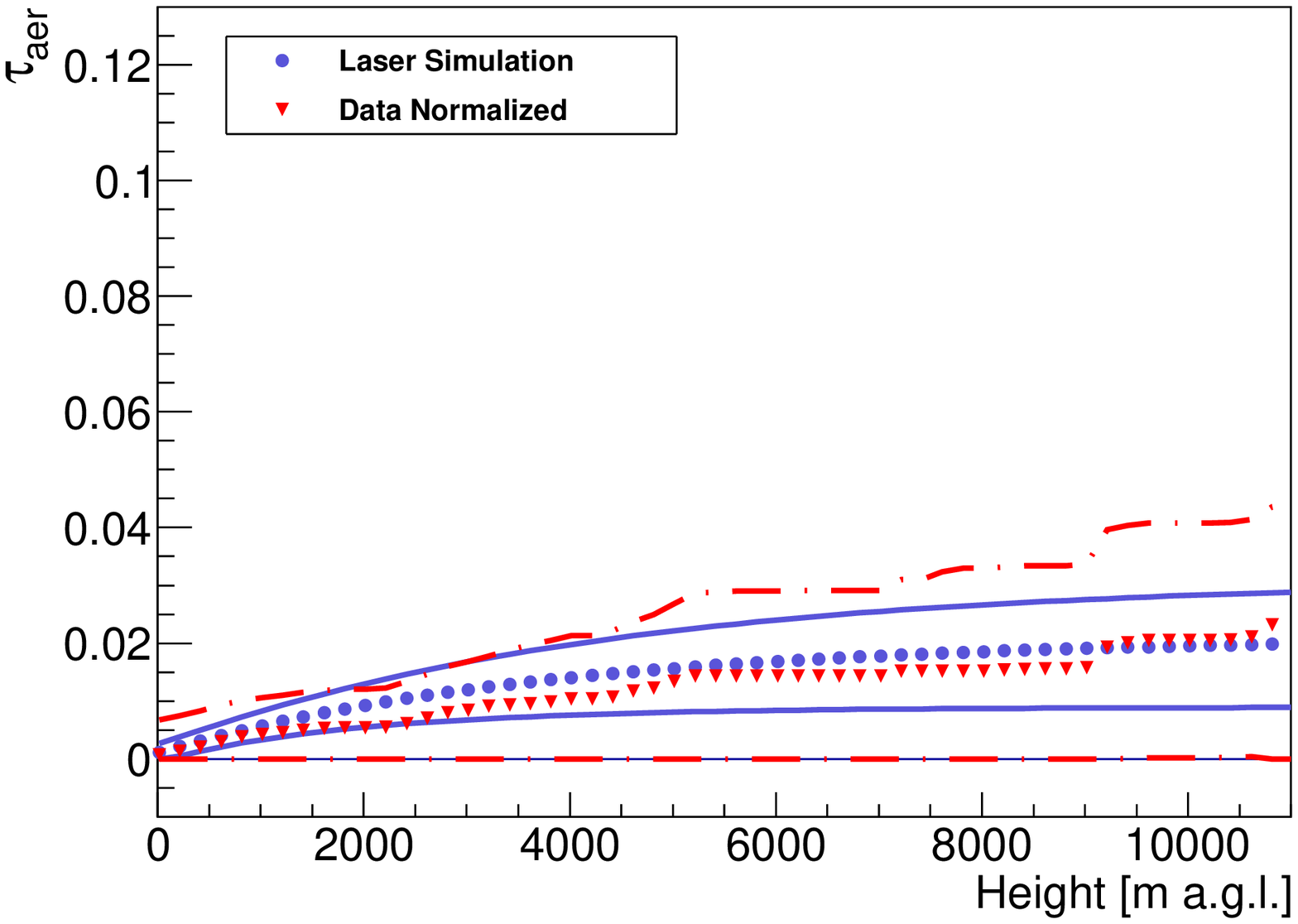}
    \caption*{(b) Low aerosol attenuation.}
  \end{minipage}
  \begin{minipage}[t]{.49\textwidth}
    \centering
    \includegraphics*[width=0.99\textwidth,clip]{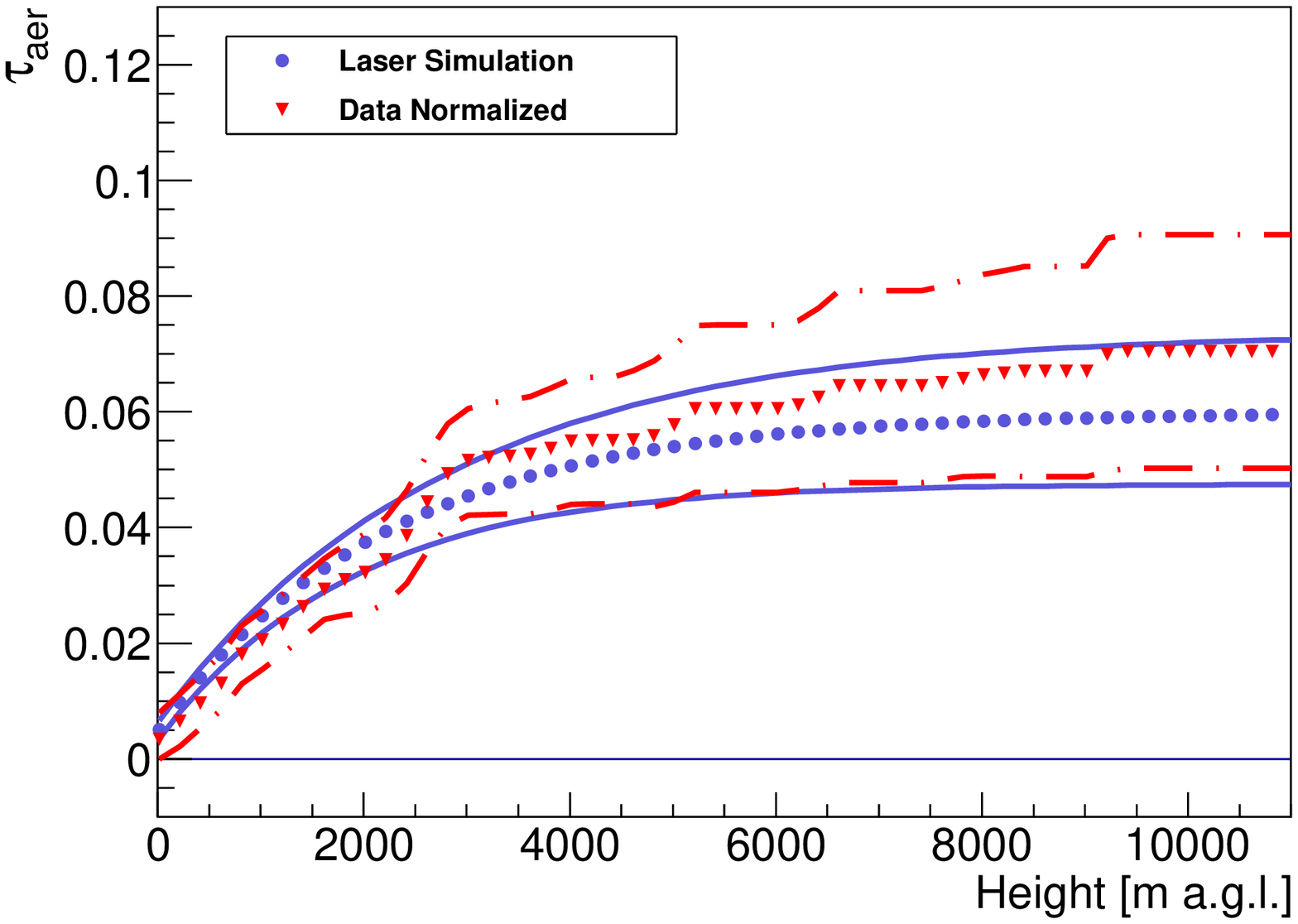}
    \caption*{(c) Average aerosol attenuation.}
  \end{minipage}
  \hfill
  \begin{minipage}[t]{.49\textwidth}
    \centering
    \includegraphics*[width=0.99\textwidth,clip]{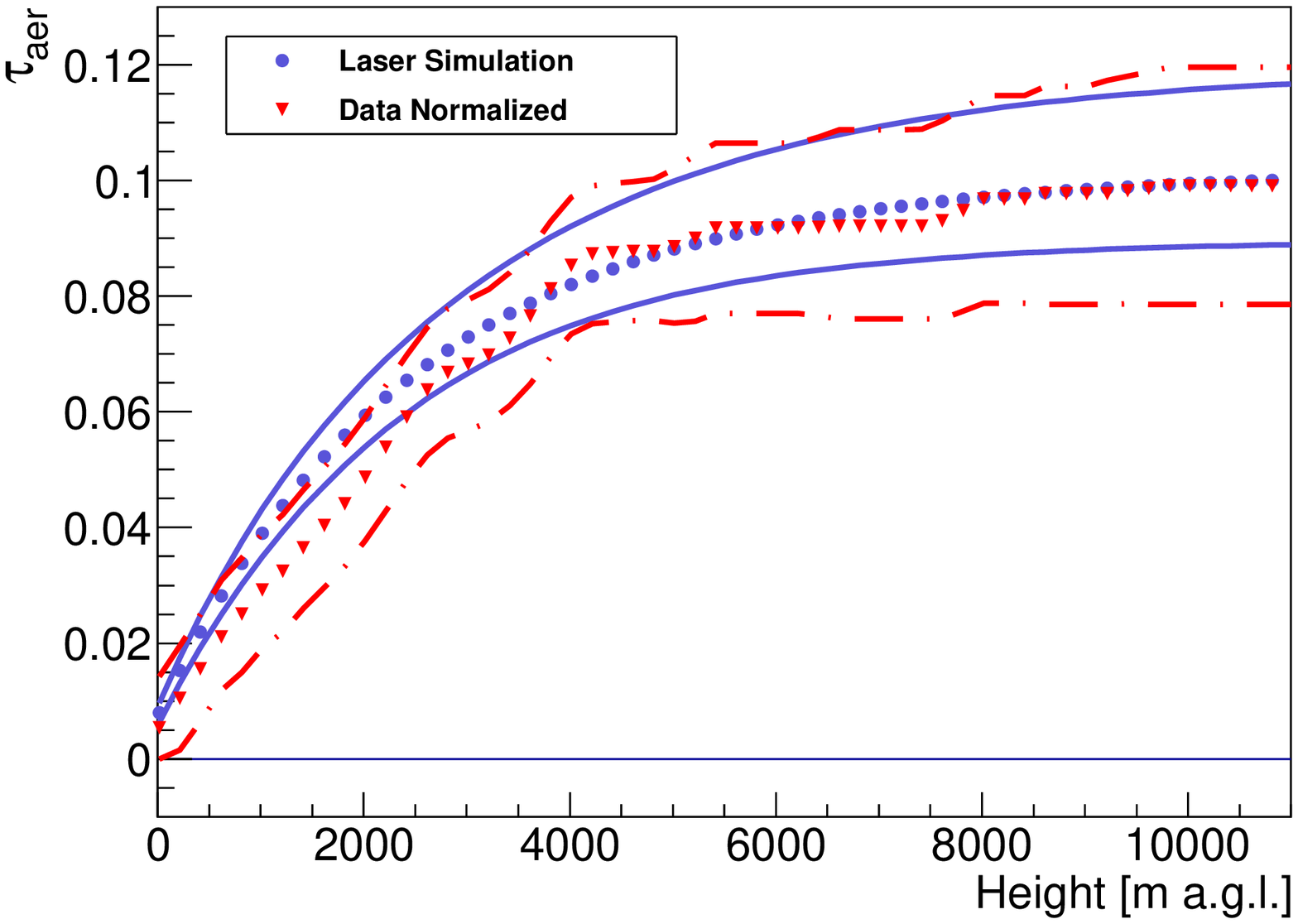}
    \caption*{(d) High aerosol attenuation.}
  \end{minipage}
  \caption{
    \label{fig:comparison}
    Correlation between $\tau_{\rm aer}$(5~km) obtained with the Laser
    Simulation and the Data Normalized procedures (a) for the year 2008
    (compatibility of results is equivalent in the other years). The dashed line
    is a diagonal indicating perfect agreement, the solid line is a fit to the
    data.
    Also shown is the vertical aerosol optical depth profile $\tau_{\rm aer}(h)$
    above ground from Laser Simulation (blue) and Data Normalized (red) analyses
    in atmospheric conditions with a low (b), average (c), and high (d) aerosol
    concentration together with the corresponding uncertainties. The laser data
    was recorded with the FD at Los Leones on July 8th, 2008 between 8~and
    9~a.m., April 4th, 2008 between 4~and 5~a.m., and January 5th, 2008 between
    3~and 4~a.m.\ local time, respectively.
  }
  \hfill
\end{figure}

Both analyses have been applied to the whole data set of CLF laser shots. A
systematic comparison of the results shows excellent agreement. Since aerosols
are concentrated in the lower part of the troposphere, we compare the total
vertical aerosol optical depth at 5~km above the FD which includes most of the
aerosols. The correlation of $\tau_{\rm aer}$(5~km) results of the Data
Normalized Analysis and the results of the Laser Simulation Analysis is shown in
Fig.~\ref{fig:comparison}.  The dashed line is a diagonal indicating perfect
agreement between the analyses.  The solid line is an actual fit to the data. It
is compatible with the diagonal.  The reliability of the parametric aerosol
model adopted and the validity of both methods can be concluded. 
In high aerosol attenuation conditions, compatible with the presence of a high Planetary Boundary Layer, 
that the Laser Simulation Analysis does not take into account, the difference between 
the measured $\tau_{\rm aer}$(5~km) is within the quoted systematic uncertainties. 
Also shown in Fig.~\ref{fig:comparison} are examples for the $\tau_{\rm aer}(h)$ profiles
estimated with the two analyses for conditions with low, average and high
aerosol attenuation, respectively.

The high compatibility of the two analyses guarantees a reliable shower
reconstruction using aerosol attenuation for the highest possible number of
hours. Nearly six years of data have been collected and analyzed (from January
2005 to September 2010). Long term results are shown in the following figures.
In the left column of Fig.~\ref{fig:vaod_distribution}, the time profile of the
vertical aerosol optical depth measured 5~km above ground using the Los Leones,
Los Morados and Coihueco FD sites is shown. The Loma Amarilla FD site is too far
from the CLF to obtain fully reliable results. The XLF is closer and will
produce aerosol attenuation measurements for Loma Amarilla in the near future.
Values of $\tau_{\rm aer}$(5~km) measured during austral winter are
systematically lower than in summer.

\begin{figure}[t]
  \begin{center}
  \begin{minipage}[t]{.49\textwidth}
    \centering
    \includegraphics*[width=0.99\textwidth,clip]{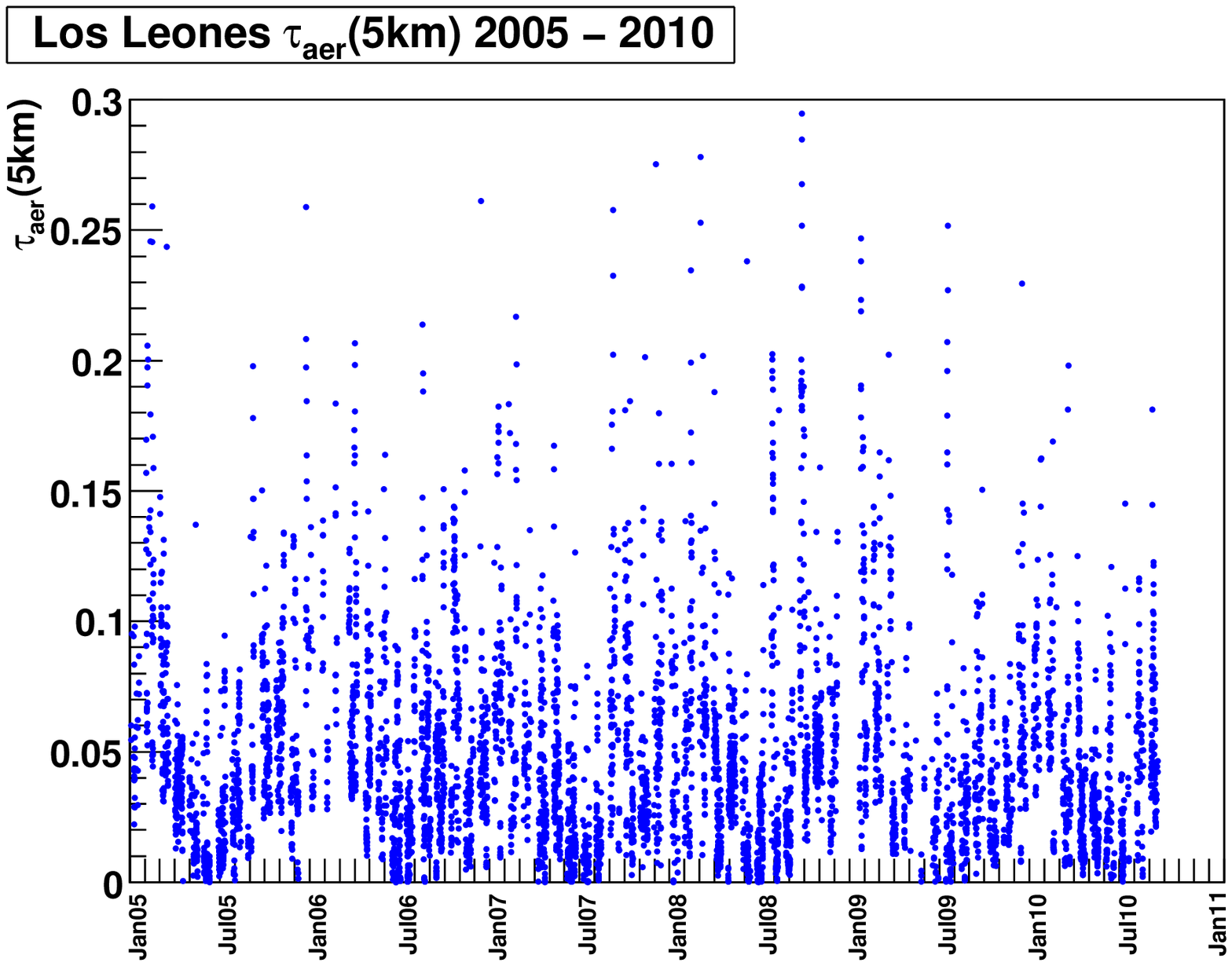}
  \end{minipage}
  \hfill
  \begin{minipage}[t]{.49\textwidth}
    \centering
    \includegraphics*[width=0.99\textwidth,clip]{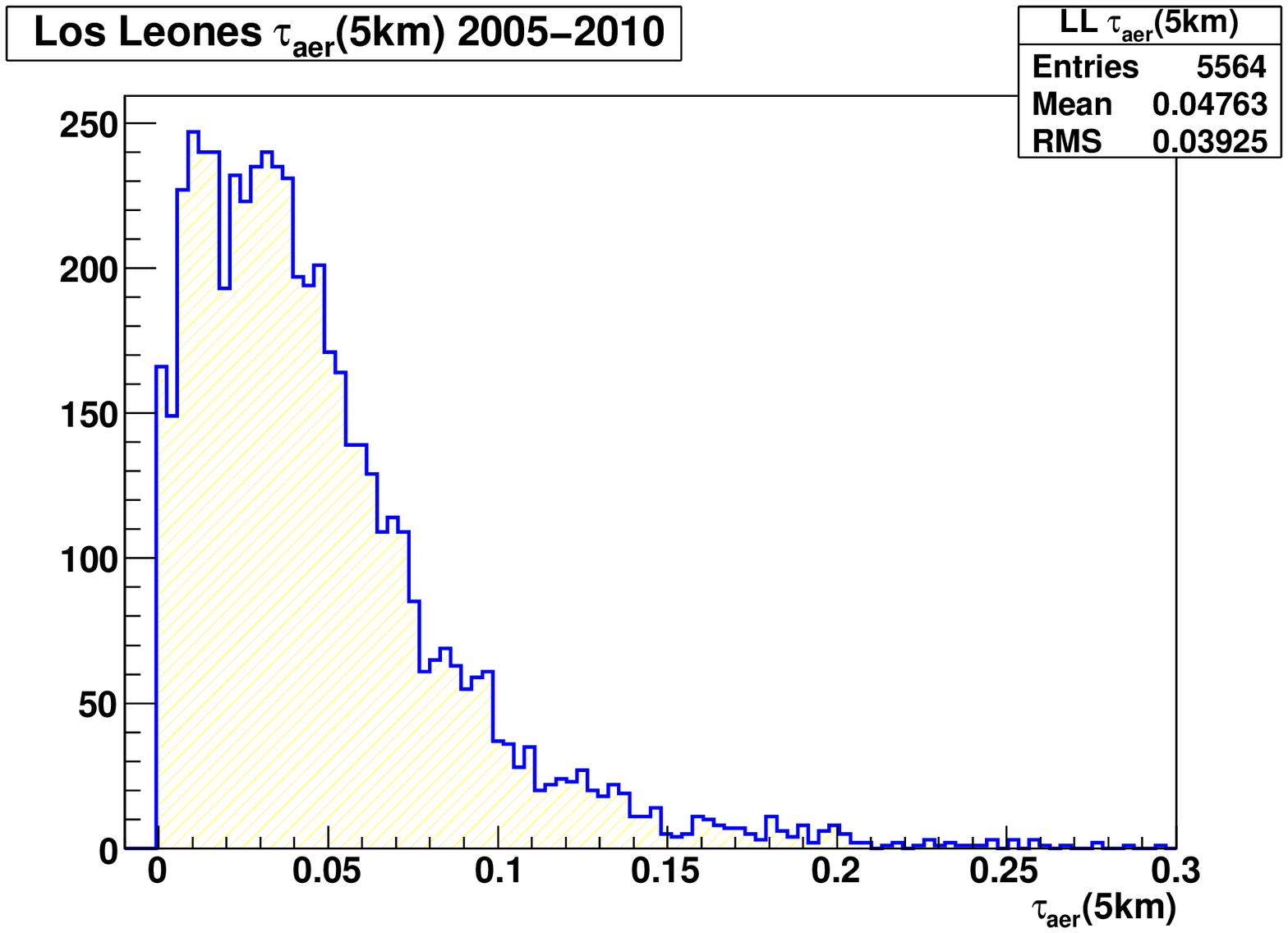}
  \end{minipage}
  \begin{minipage}[t]{.49\textwidth}
    \centering
    \includegraphics*[width=0.99\textwidth,clip]{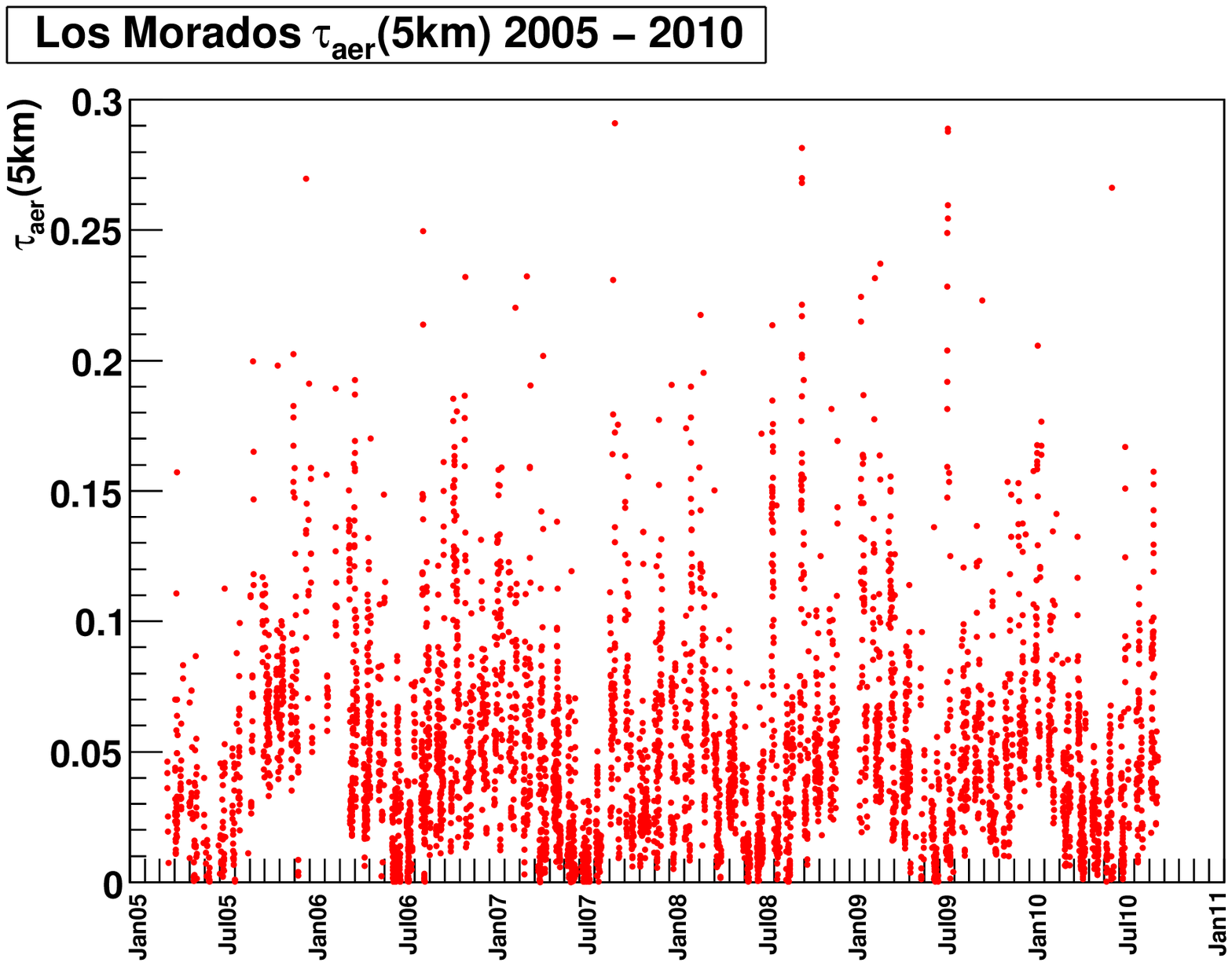}
  \end{minipage}
  \hfill
  \begin{minipage}[t]{.49\textwidth}
    \centering
    \includegraphics*[width=0.99\textwidth,clip]{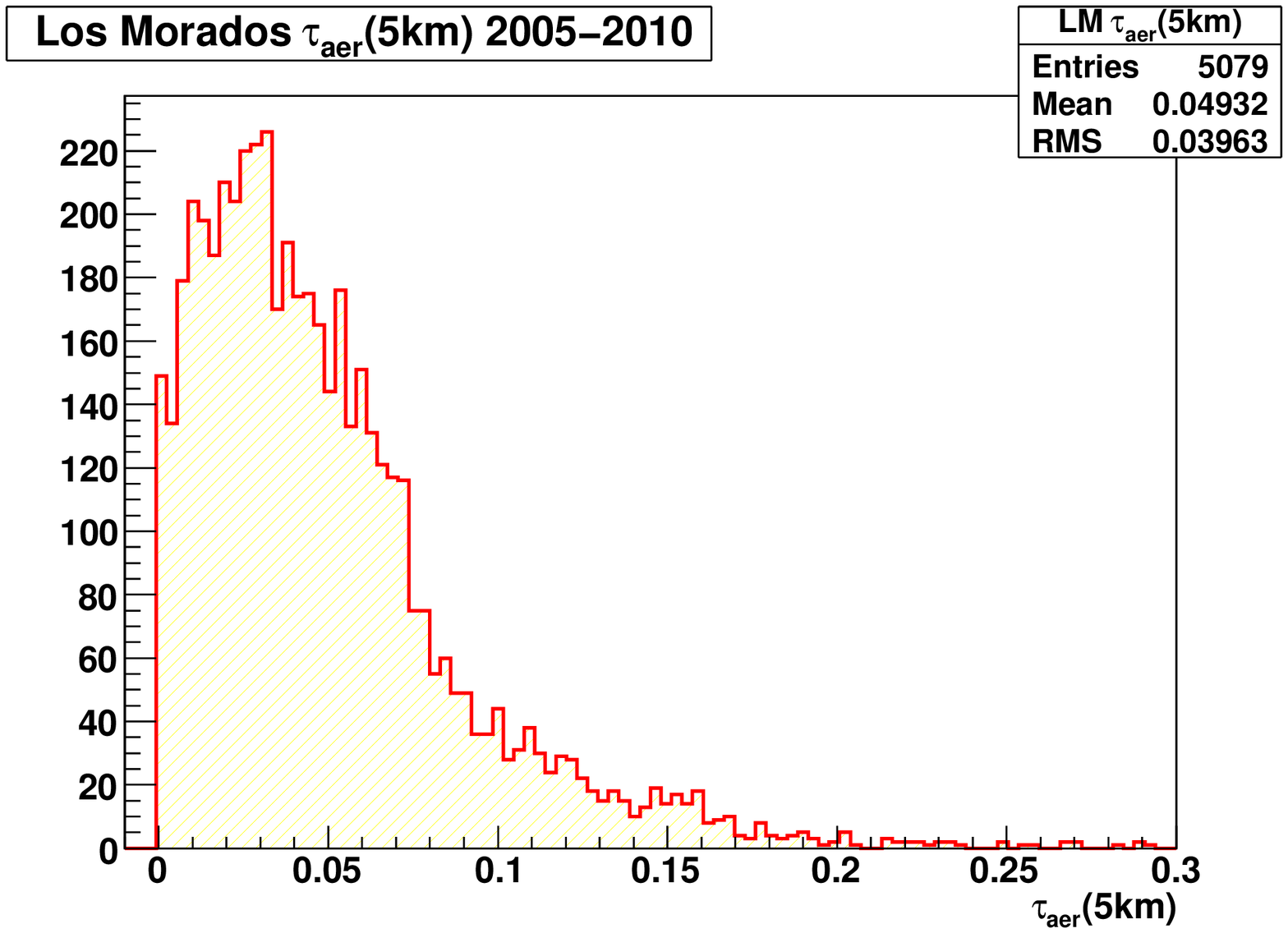}
  \end{minipage}
  \begin{minipage}[t]{.49\textwidth}
    \centering
    \includegraphics*[width=0.99\textwidth,clip]{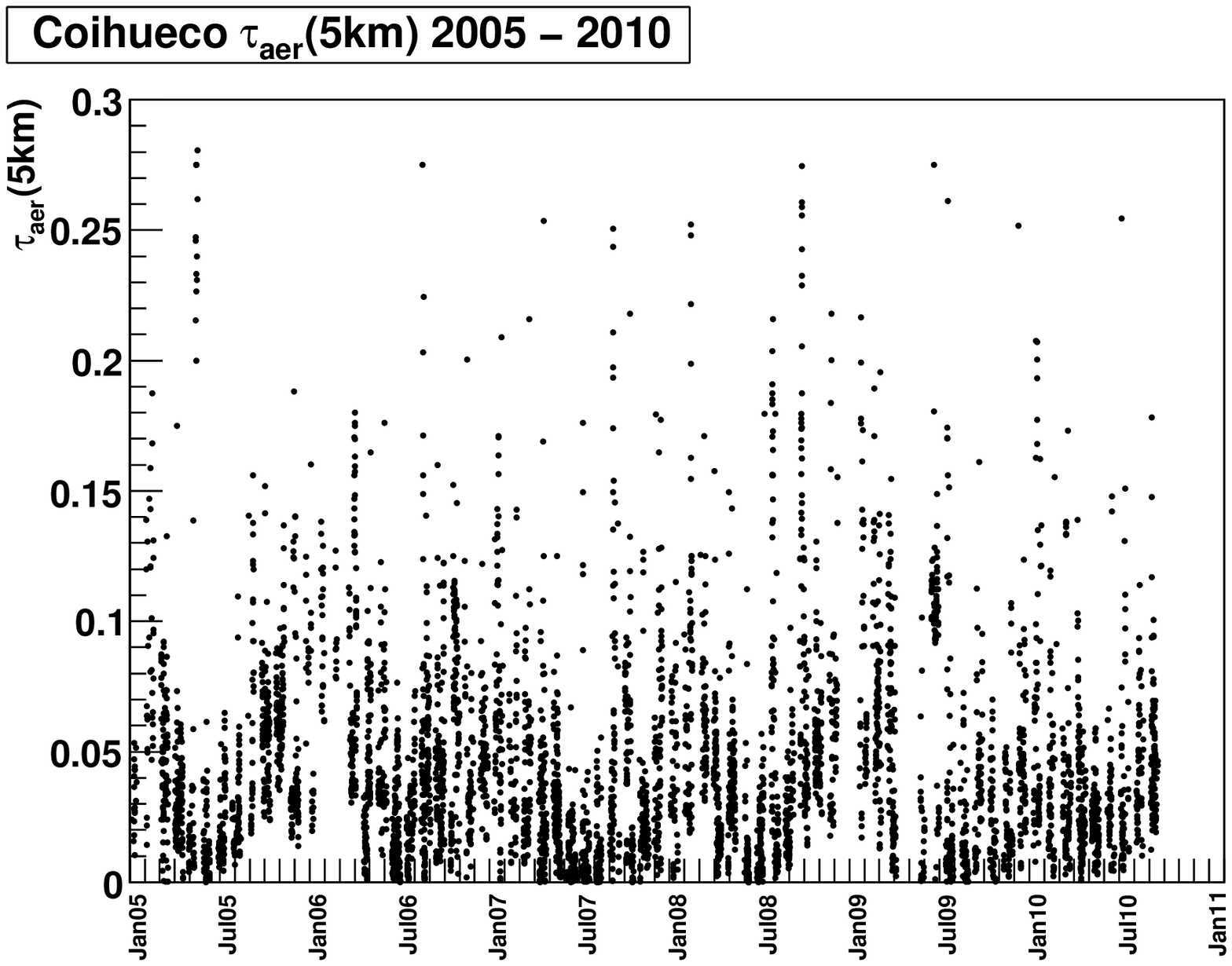}
  \end{minipage}
  \hfill
  \begin{minipage}[t]{.49\textwidth}
    \centering
    \includegraphics*[width=0.99\textwidth,clip]{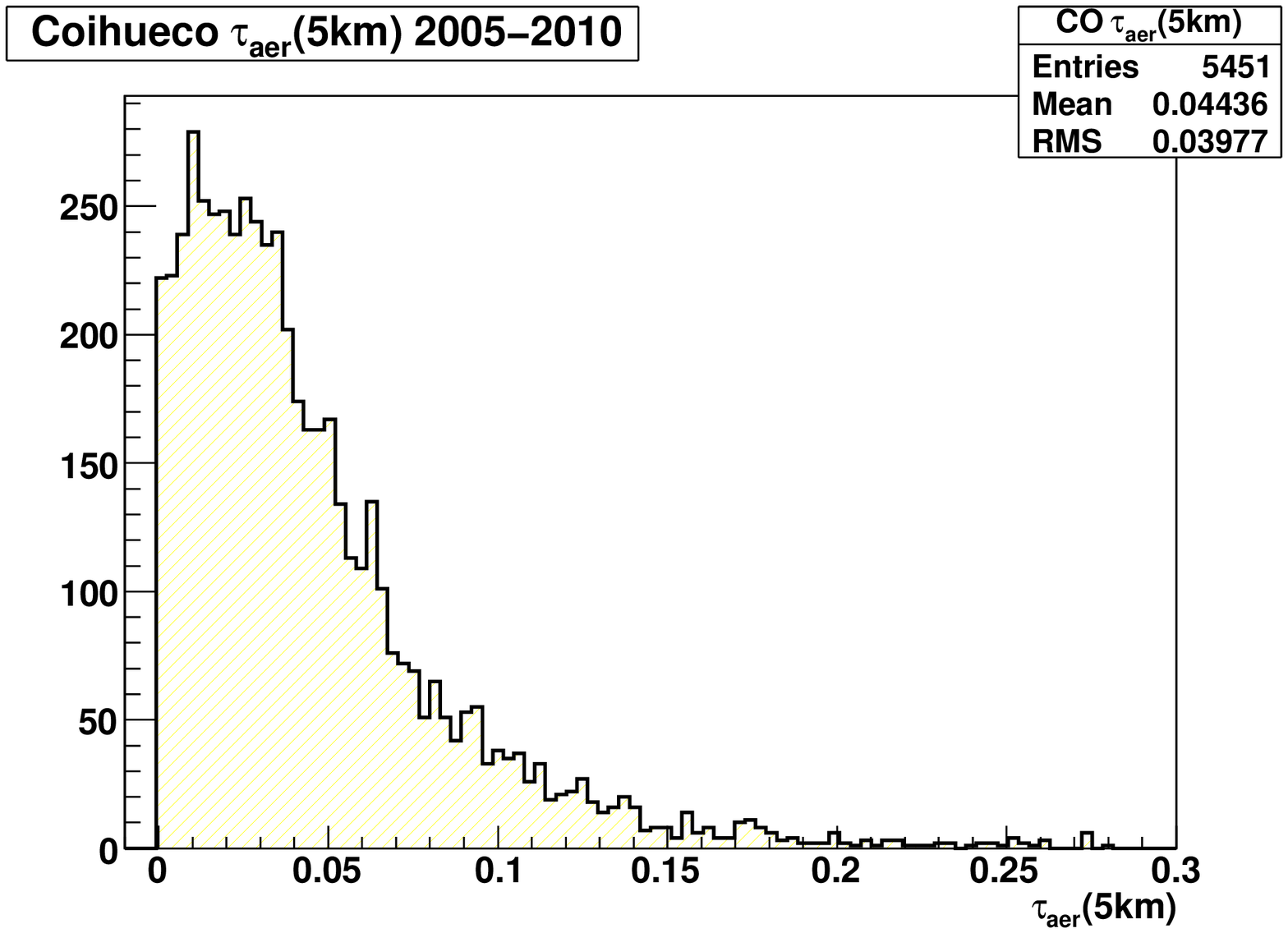}
  \end{minipage}
  \caption{
    \label{fig:vaod_distribution}
    Vertical aerosol optical depth $\tau_{\rm aer}$ 5~km above the ground,
    measured with the Los Leones (top), Los Morados (middle) and Coihueco
    (bottom) FD sites. Left column: Hourly measurements of $\tau_{\rm aer}$ versus
    time. Right column: Distribution of hourly measurements of $\tau_{\rm aer}$.
    Average values are very similar.
  }
  \end{center}
\end{figure}

In the right column of Fig.~\ref{fig:vaod_distribution}, the $\tau_{\rm
aer}$(5~km) distribution over six years is shown for aerosol attenuation
measurements using the FD sites at Los Leones, Los Morados and Coihueco. More
than 5000~hours of aerosol profiles have been measured with each FD.  The
average $\tau_{\rm aer}$(5~km) measured with different FD sites are compatible.
The average value measured above Coihueco is slightly smaller due to the higher
position ($\sim$ 300~m) of the Coihueco FD site with respect to Los Leones and
Los Morados.

%% file: conclusions.tex
\section{Conclusions}
\label{sec:conclusions}

Aerosols cause the largest time-varying corrections applied during the
reconstruction of extensive air showers measured with the fluorescence
technique. They are highly variable on a time scale of one hour. Neglecting the
aerosol attenuation leads to a bias in the energy reconstruction of air showers
by 8~to 25\% in the energy range measured by the \pao. This includes a tail of
7\% of all showers with an energy correction larger than 30\%.

To determine the vertical aerosol optical depth profiles for the \pao, vertical
laser shots from a Central Laser Facility in the center of the SD array are
analyzed. The Central Laser Facility fires 50~vertical shots every 15~minutes
during the FD data acquisition, covering the whole FD data taking period. Two
methods were developed to analyze the CLF laser shots. The Data Normalized
method compares the measured laser light profile to a reference clear night, the
Laser Simulation method compares the measured profile with a set of simulated
profiles. In addition, the minimum cloud heights over the central part of the
array are extracted from the laser data. The two methods are compared and a very
good agreement was found. Nearly six years of data have been analyzed with both
methods (from January 2005 to September 2010). In air shower reconstructions,
mainly the results of the Data Normalized method are used. The data from the
Laser Simulation method is used to fill holes in the data set where the Data
Normalized method is not able to produce a result.

%% file: acknowledgments.tex
The successful installation, commissioning, and operation of the Pierre Auger Observatory including the Central Laser Facility
would not have been possible without the strong commitment and effort
from the technical and administrative staff in Malarg\"ue.

We are very grateful to the following agencies and organizations for financial support:
Comisi\'on Nacional de Energ\'ia At\'omica,
Fundaci\'on Antorchas,
Gobierno De La Provincia de Mendoza,
Municipalidad de Malarg\"ue,
NDM Holdings and Valle Las Le\~nas, in gratitude for their continuing
cooperation over land access, Argentina;
the Australian Research Council;
Conselho Nacional de Desenvolvimento Cient\'ifico e Tecnol\'ogico (CNPq),
Financiadora de Estudos e Projetos (FINEP),
Funda\c{c}\~ao de Amparo \`a Pesquisa do Estado de Rio de Janeiro (FAPERJ),
Funda\c{c}\~ao de Amparo \`a Pesquisa do Estado de S\~ao Paulo (FAPESP),
Minist\'erio de Ci\^{e}ncia e Tecnologia (MCT), Brazil;
AVCR AV0Z10100502 and AV0Z10100522,
GAAV KJB100100904,
MSMT-CR LA08016, LC527, 1M06002, MEB111003, and MSM0021620859, Czech Republic;
Centre de Calcul IN2P3/CNRS,
Centre National de la Recherche Scientifique (CNRS),
Conseil R\'egional Ile-de-France,
D\'epartement  Physique Nucl\'eaire et Corpusculaire (PNC-IN2P3/CNRS),
D\'epartement Sciences de l'Univers (SDU-INSU/CNRS), France;
Bundesministerium f\"ur Bildung und Forschung (BMBF),
Deutsche Forschungsgemeinschaft (DFG),
Finanzministerium Baden-W\"urttemberg,
Helm\-holtz-Ge\-mein\-schaft Deut\-scher For\-schungs\-zen\-tren (HGF),
Mi\-nis\-te\-ri\-um f\"ur Wis\-sen\-schaft und For\-schung, Nord\-rhein-West\-falen,
Mi\-nis\-te\-ri\-um f\"ur Wis\-sen\-schaft, For\-schung und Kunst, Ba\-den-W\"urt\-tem\-berg, Germany;
Istituto Nazionale di Fisica Nucleare (INFN),
Ministero dell'Istruzione, dell'U\-ni\-ver\-si\-t\`a e della Ricerca (MIUR), Italy;
Consejo Nacional de Ciencia y Tecnolog\'ia (CONACYT), Mexico;
Ministerie van Onderwijs, Cultuur en Wetenschap,
Nederlandse Organisatie voor Wetenschappelijk Onderzoek (NWO),
Stichting voor Fundamenteel Onderzoek der Materie (FOM), Netherlands;
Ministry of Science and Higher Education,
Grant Nos. N N202 200239 and N N202 207238, Poland;
Funda\c{c}\~ao para a Ci\^{e}ncia e a Tecnologia, Portugal;
Ministry for Higher Education, Science, and Technology,
Slovenian Research Agency, Slovenia;
Comunidad de Madrid,
Consejer\'ia de Educaci\'on de la Comunidad de Castilla La Mancha,
FEDER funds,
Ministerio de Ciencia e Innovaci\'on and Consolider-Ingenio 2010 (CPAN),
Xunta de Galicia, Spain;
Science and Technology Facilities Council, United Kingdom;
Department of Energy, Contract Nos. DE-AC02-07CH11359, DE-FR02-04ER41300,
National Science Foundation, Grant Nos. 0450696, 0855680,
The Grainger Foundation USA;
NAFOSTED, Vietnam;
ALFA-EC / HELEN,
European Union 6th Framework Program,
Grant No. MEIF-CT-2005-025057,
European Union 7th Framework Program, Grant No. PIEF-GA-2008-220240,
and UNESCO.

%% file: CLFpaper.bbl
\providecommand{\href}[2]{#2}\begingroup\raggedright\begin{thebibliography}{10}

\bibitem{auger}
The Pierre Auger Collaboration, J.~Abraham et~al., {\it {Properties and
  performance of the prototype instrument for the Pierre Auger Observatory}},
  {\em Nucl. Instr. Meth.} {\bf A523} (2004) 50--95.

\bibitem{FD}
The Pierre Auger Collaboration, J.~Abraham et~al., {\it {The Fluorescence
  Detector of the Pierre Auger Observatory}},  {\em Nucl. Instr. Meth.} {\bf
  A620} (2010) 227--251, [\href{http://xxx.lanl.gov/abs/0907.4282}{{\tt
  arXiv:0907.4282}}].

\bibitem{segev_atmo}
The Pierre Auger Collaboration, J.~Abraham et~al., {\it {A Study of the Effect
  of Molecular and Aerosol Conditions in the Atmosphere on Air Fluorescence
  Measurements at the Pierre Auger Observatory}},  {\em Astropart. Phys.} {\bf
  33} (2010) 108--129, [\href{http://xxx.lanl.gov/abs/1002.0366}{{\tt
  arXiv:1002.0366}}].

\bibitem{clfjinst}
B.~Fick et~al., {\it {The Central Laser Facility at the Pierre Auger
  Observatory}},  {\em JINST} {\bf 1} (2006) P11003,
  [\href{http://xxx.lanl.gov/abs/astro-ph/0507334}{{\tt astro-ph/0507334}}].

\bibitem{lidar_paper}
S.~Y. BenZvi et~al., {\it {The Lidar System of the Pierre Auger Observatory}},
  {\em Nucl. Instr. Meth.} {\bf A574} (2007) 171--184,
  [\href{http://xxx.lanl.gov/abs/astro-ph/0609063}{{\tt astro-ph/0609063}}].

\bibitem{karim_icrc}
K.~Louedec for the Pierre Auger Collaboration, {\it {Atmospheric Monitoring at
  the Pierre Auger Observatory -- Status and Update}},  in {\em Proc. 32nd
  ICRC}, vol.~2, (Beijing, China), pp.~63--66, 2011,
  [\href{http://xxx.lanl.gov/abs/1107.4806}{{\tt arXiv:1107.4806}}].

\bibitem{apf}
S.~Y. BenZvi et~al., {\it {Measurement of the aerosol phase function at the
  Pierre Auger Observatory}},  {\em Astropart. Phys.} {\bf 28} (2007) 312--320,
  [\href{http://xxx.lanl.gov/abs/0704.0303}{{\tt arXiv:0704.0303}}].

\bibitem{king_factor}
L.~V. King, {\it On the complex anisotropic molecule in relation to the
  dispersion and scattering of light},  {\em Proc. R. Soc. London Ser. A} {\bf
  104} (1923) 333--357.

\bibitem{epj_atmo}
B.~Keilhauer and M.~Will, {\it {Description of Atmospheric Conditions at the
  Pierre Auger Observatory Using Meteorological Measurements and Models}},
  {\em Eur. Phys. J. Plus} {\bf 127} (2012) 96--105,
  [\href{http://xxx.lanl.gov/abs/1208.5417}{{\tt arXiv:1208.5417}}].

\bibitem{gdas_paper}
The Pierre Auger Collaboration, P.~Abreu et~al., {\it {Data from the Global
  Data Assimilation System (GDAS) for the Pierre Auger Observatory}},  {\em
  Astropart. Phys.} {\bf 35} (2012) 591--607,
  [\href{http://xxx.lanl.gov/abs/1201.2276}{{\tt arXiv:1201.2276}}].

\bibitem{maria_aerosol}
M.~I. Micheletti et~al., {\it {Elemental analysis of aerosols collected at the
  Pierre Auger Cosmic Ray Observatory with PIXE technique complemented with
  SEM/EDX}},  {\em Nucl. Instr. Meth.} {\bf B288} (2012) 10--17.

\bibitem{airfly}
The AIRFLY Collaboration, M.~Ave et~al., {\it {Spectrally resolved pressure
  dependence measurements of air fluorescence emission with AIRFLY}},  {\em
  Nucl. Inst. Meth.} {\bf A597} (2008) 41--45.

\bibitem{wiencke_upgrade}
L.~Wiencke et~al. for the Pierre Auger Collaboration, {\it {Atmospheric ``Super
  Test Beam'' for the Pierre Auger Observatory}},  in {\em Proc. 32nd ICRC},
  vol.~3, (Beijing, China), pp.~141--144, 2011,
  [\href{http://xxx.lanl.gov/abs/1107.4806}{{\tt arXiv:1107.4806}}].

\bibitem{Abbasi}
The High Resolution Fly's Eye Collaboration (HiRes), R.~U. Abbasi et~al., {\it
  {Techniques for measuring atmospheric aerosols at the High Resolution Fly's
  Eye experiment}},  {\em Astropart. Phys.} {\bf 25} (2006) 74--83,
  [\href{http://xxx.lanl.gov/abs/astro-ph/0512423}{{\tt astro-ph/0512423}}].

\end{thebibliography}\endgroup
